\input harvmac 
\input epsf.tex

\overfullrule=0mm

\newcount\figno
\figno=0
\def\fig#1#2#3{
\par\begingroup\parindent=0pt\leftskip=1cm\rightskip=1cm\parindent=0pt
\baselineskip=11pt
\global\advance\figno by 1
\midinsert
\epsfxsize=#3
\centerline{\epsfbox{#2}}
\vskip 12pt
{\bf Fig. \the\figno:} #1\par
\endinsert\endgroup\par
}
\def\figlabel#1{\xdef#1{\the\figno}}
\def\encadremath#1{\vbox{\hrule\hbox{\vrule\kern8pt\vbox{\kern8pt
\hbox{$\displaystyle #1$}\kern8pt}
\kern8pt\vrule}\hrule}}

\def\wrt{with respect to\ }

\def\IR{\relax{\rm I\kern-.18em R}}
\font\cmss=cmss10 \font\cmsss=cmss10 at 7pt

\font\cmss=cmss10 \font\cmsss=cmss10 at 7pt
\def\IZ{\relax\ifmmode\mathchoice
{\hbox{\cmss Z\kern-.4em Z}}{\hbox{\cmss Z\kern-.4em Z}}
{\lower.9pt\hbox{\cmsss Z\kern-.4em Z}}
{\lower1.2pt\hbox{\cmsss Z\kern-.4em Z}}\else{\cmss Z\kern-.4em Z}\fi}
\def\IN{\relax{\rm I\kern-.18em N}}


\Title{\vbox{\hsize=3.truecm \hbox{SPhT/02-007}}}
{{\vbox {
\bigskip
\centerline{Critical and Tricritical Hard Objects}
\centerline{on Bicolorable Random Lattices:}
\centerline{Exact Solutions}
}}}
\bigskip
\centerline{J. Bouttier\foot{bouttier@spht.saclay.cea.fr}, 
P. Di Francesco\foot{philippe@spht.saclay.cea.fr} and
E. Guitter\foot{guitter@spht.saclay.cea.fr}}
\medskip
\centerline{ \it CEA-Saclay, Service de Physique Th\'eorique,}
\centerline{ \it F-91191 Gif sur Yvette Cedex, France}
\bigskip
\noindent We address the general problem of hard objects on random lattices,
and emphasize the crucial role played by the colorability of the lattices to
ensure the existence of a crystallization transition. We first solve explicitly
the naive (colorless) random-lattice version of the hard-square model and find
that the only matter critical point is the non-unitary Lee-Yang edge
singularity. We then show how to restore the crystallization transition of the
hard-square model by considering the same model on bicolored random lattices.
Solving this model exactly, we show moreover that the crystallization
transition point lies in the universality class of the Ising model coupled to
2D quantum gravity. We finally extend our analysis to a new two-particle
exclusion model, whose regular lattice version involves hard squares of two
different sizes. The exact solution of this model on bicolorable random
lattices displays a phase diagram with two (continuous and discontinuous)
crystallization transition lines meeting at a higher order critical point,
in the universality class of the tricritical Ising model coupled to 2D quantum
gravity.

\Date{01/02}

\nref\BAXHH{R. J. Baxter, {\it Hard Hexagons: Exact Solution}, J. Phys. {\bf A 13}
(1980) L61-L70; R. J. Baxter and S.K. Tsang, {\it Entropy of Hard Hexagons},
J. Phys. {\bf A 13} (1980) 1023-1030; see also
R. J. Baxter, {\it Exactly Solved Models in Statistical Mechanics},
Academic Press, London (1984).}
\nref\CARTWO{J. Cardy, {\it Conformal Invariance and the Yang-Lee Edge
Singularity in Two Dimensions}, 
Phys. Rev. Lett. {\bf 54}, No. 13 (1985) 1354-1356.}
\nref\DSZ{P. Di Francesco, H. Saleur and J.-B. Zuber, 
{\it Generalized Coulomb Gas Formalism for Two-dimensional Critical
Models based on SU(2) Coset Construction}, Nucl. Phys. {\bf B300
[FS]} (1988) 393-432.}
\nref\TAKA{H. Takasaki, T. Nishino and Y. Hieida, {\it Phase Diagram 
of a 2D Vertex Model}, J. Phys. Soc. Japan Vol. {\bf 70} (2001) 1429-1430, 
preprint cond-mat/0012490.}
\nref\BAX{R. J. Baxter, {\it Planar Lattice Gases with Nearest-neighbour
Exclusion}, Annals of Combin. No. {\bf 3} (1999) 191-203 preprint cond-mat/9811264.}
\nref\BAXHS{R. J. Baxter, I. G. Enting and S.K. Tsang, {\it Hard Square Lattice Gas},
J. Stat. Phys. {\bf 22} (1980) 465-489.}
\nref\GF{D. Gaunt and M. Fisher, 
{\it Hard-Sphere Lattice Gases.I.Plane-Square Lattice}, J. Chem. Phys. 
{\bf 43} (1965) 2840-2863.}
\nref\RUN{L. Runnels, L. Combs and J. Salvant, {\it Exact Finite 
Methods of Lattice Statistics. II. Honeycomb-Lattice Gas of Hard Molecules}, 
J. Chem. Phys. {\bf 47} (1967) 4015-4020.}
\nref\DGZ{P. Di Francesco, P. Ginsparg
and J. Zinn--Justin, {\it 2D Gravity and Random Matrices},
Physics Reports {\bf 254} (1995) 1-131.}
\nref\EY{B. Eynard, {\it Random Matrices}, Saclay Lecture Notes (2000),
\hfill\break 
http://www-spht.cea.fr/lectures\_notes.shtml }
\nref\KF{D. Kurze and M. Fisher, {\it Yang-Lee Edge Singularities
at High Temperatures}, Phys. Rev. {\bf B20} (1979) 2785-2796.}
\nref\INV{P. Di Francesco and  E. Guitter, {\it Critical and Multicritical 
Semi-Random (1+d)-Dimensional Lattices and Hard Objects
in d Dimensions}, preprint cond-mat/0104383 (2001), 
to appear in J. Phys. {\bf A} (2002).}
\nref\HDIM{M. Staudacher, {\it The Yang-Lee Edge Singularity on a 
Dynamical Planar Random Surface}, Nucl. Phys. {\bf B336} (1990) 349-362.}
\nref\BIPZ{E. Br\'ezin, C. Itzykson, G. Parisi and J.-B. Zuber, {\it Planar
Diagrams}, Comm. Math. Phys. {\bf 59} (1978) 35-51.}
\nref\CK{L. Chekhov and C. Kristjansen, {\it Hermitian Matrix Model with 
Plaquette Interaction}, Nucl.Phys. {\bf B479} (1996) 683-696.}
\nref\TUT{W. Tutte, {\it A Census of Planar Maps}, Canad. Jour. of Math. 
{\bf 15} (1963) 249-271.}
\nref\DEG{P. Di Francesco, B. Eynard and E. Guitter, 
{\it Coloring Random Triangulations},
Nucl. Phys. {\bf B516 [FS]} (1998) 543-587.}
\nref\BEG{M. Blume, V. Emery and R. Griffiths, {\it Ising Model for the 
$\lambda$ Transition and Phase Separation in He$^3$-He$^4$ Mixtures},
Phys. Rev. {\bf A4} (1971) 1071-1077.}
\nref\DG{I. Lawrie and S. Sarbach, {\it Theory of Tricritical Points},
in {\it Phase Transitions and Critical Phenomena}, vol. 9, C. Domb and J. Lebowitz eds., 
Academic Press, London (1984).}
\nref\FZ{V.A. Fateev and A.B. Zamolodchikov, {\it Selfdual Solutions of the Star-triangle
Relations in $\IZ_N$ Models}, Phys. Lett. {\bf A92} (1982) 37-39;
A.B. Zamolodchikov and V.A. Fateev, {\it Nonlocal (Parafermion) Currents in
Two-dimensional Conformal Quantum Field Theory and Self-dual Critical Points
in $\IZ_n$-symmetric Statistical Systems}, Sov. Phys. J.E.T.P. {\bf 62} (1985) 215-225.} 
\nref\GKN{E. Guitter, C. Kristjansen and J.L. Nielsen {\it Hamiltonian
Cycles on Random Eulerian Triangulations} Nucl. Phys {\bf B546[FS]} (1999)
731-750.}
\nref\DGK{P. Di Francesco, E. Guitter and C. Kristjansen {\it Fully Packed
O(n=1) Model on Random Eulerian Triangulations} Nucl. Phys {\bf B549[FS]} 
(1999) 657-667.}
\nref\KPZ{V.G. Knizhnik, A.M. Polyakov and A.B. Zamolodchikov, {\it Fractal Structure of
2D Quantum Gravity}, Mod. Phys. Lett.
{\bf A3} (1988) 819-826; F. David, {\it Conformal Field Theories Coupled to 2D Gravity in the
Conformal Gauge}, Mod. Phys. Lett. {\bf A3} (1988) 1651-1656; J.
Distler and H. Kawai, {\it Conformal Field Theory and 2D Quantum Gravity}, 
Nucl. Phys. {\bf B321} (1989) 509-527.}

\newsec{Introduction}

In lattice statistical mechanics, universality classes usually do not depend on  
the lattice over which the model is defined, but only on the symmetries of
the interactions. 
The situation however becomes more subtle when the symmetries of the interactions
themselves strongly depend on the properties of the underlying lattice. 
A famous example of such behavior is provided by the general problem
of ``hard" objects on two-dimensional lattices. In these models, each site
of the lattice may be in two states, occupied or empty, but if a site
is occupied, then necessarily its nearest neighbors must be empty. The 
models are further defined by attaching an activity $z$ per occupied vertex.
For regular lattices in two dimensions,
an exact solution for the thermodynamics of the model exists so far only in the case of
the triangular lattice (hard hexagon model, solved by Baxter \BAXHH). 
In this case, two critical points were found at values 
$z_\pm=\big({1\pm\sqrt{5}\over 2}\big)^5$, $z_-<0$, $z_+>0$,
respectively governed by the Lee-Yang edge singularity in 2D [\xref\CARTWO,\xref\DSZ] 
(non-unitary Conformal Field Theory (CFT) with central
charge $c(2,5)=-22/5$) and the critical three-state Potts model (unitary
CFT with central charge $c(5,6)=4/5$).
Despite recent progress [\xref\TAKA,\xref\BAX], the hexagonal lattice and square
lattice cases remain elusive \BAXHS. Numerical evidence however seems
to indicate that they still have two critical points $z_-<0$ and $z_+>0$,
and while the first still corresponds to the Lee-Yang edge singularity, the other one
displays the exponents of the critical Ising model \TAKA\ (CFT with $c(3,4)=1/2$).
The difference in universality class at $z_+$ for the triangular lattice 
on one hand and the hexagonal or square lattice on the other may be simply
understood from the symmetries of the lattices. Indeed, this critical point
corresponds in all cases to a crystallization transition, where the
hard particles occupy preferentially a particular sublattice of the lattice
at hand. The common feature of the square and hexagonal lattice is their
vertex-bicolorability (bipartite nature) which naturally defines two 
equivalent mutually excluding sublattices corresponding to two 
possible symmetric crystalline groundstates. The triangular lattice on the other hand
is not vertex-bicolorable, but vertex-tricolorable instead, which allows   
to define three equivalent mutually excluding sublattices corresponding to 
three possible symmetric crystalline states.
These two- or three-fold symmetries give rise naturally to critical points
with $\IZ_2$ (Ising) or $\IZ_3$ (three-state Potts) symmetries.

The purpose of this note is to study similar hard particle models
on random lattices such as those used to generate discrete models
of 2D quantum gravity. The aim of this study is twofold:
(1) to check the crucial role played by the vertex-colorability of the underlying
lattices in determining the physical behavior of the models and
(2) to give a ``gravitational" proof that the crystallization transition point $z_+$
indeed lies in the critical Ising universality class 
in the case of hard particles on bicolorable lattices. 

More precisely, in this paper, we first solve explicitly the problem of hard particles 
on arbitrary random lattices and show that in the absence of any colorability
constraint (Euclidean random surfaces), only the
Lee-Yang critical point survives at some negative value
$z=z_-$. 
We then solve the same model on so-called Eulerian random surfaces, i.e. random lattices
for which we impose vertex-bicolorability, and find that the crystallization
transition point $z=z_+$ is restored in this case, and that it belongs to
the universality class of the critical Ising model coupled to 2D quantum gravity.  

We next extend our analysis to higher order critical points and show how to recover
the tricritical Ising universality class (CFT with central charge
$c(4,5)=7/10$) by considering a {\it two-particle}
exclusion model on vertex-bicolorable lattices. We again give a 
gravitational proof of this fact by explicitly solving the model on 
random vertex-bicolorable lattices.

The paper is organized as follows.
In Sect.\ 2 we study the model of hard particles and first recall a few known facts
on its regular honeycomb and square lattice version (Sect.\ 2.1). We then solve
the model in Sect.\ 2.2 on arbitrary random planar tetravalent lattices, by use
of a two-matrix integral. In addition to the activity $z$ per particle,
the latter include an extra weight $g$ per vertex (occupied or not). We derive
explicitly the gravitational critical line $g=g_c(z)$ selecting arbitrarily large 
lattices, allowing
to reach the interesting thermodynamic behavior of the model. On this line,
we find a unique critical point of the matter system at some $z=z_-$, which
we identify as the Lee-Yang edge singularity. In Sect.\ 2.3,    
we introduce a four-matrix model describing the same problem now on random
vertex-bicolorable lattices. We solve this model exactly in the case of
trivalent such lattices and obtain in particular the new gravitational
critical line $g=g_c(z)$. Along this line, we now find two matter critical
points at some values $z=z_-<0$ and $z=z_+>0$. The first one is still identified
as the Lee-Yang edge singularity, while the new one is identified as
a crystallization transition point in the universality class of the critical Ising
model. 
Sect.\ 3 is devoted to the study of a more sophisticated two-particle exclusion
model in which we allow sites to be occupied by single particles (with activity $z_1$
per particle)
or pairs of particles (with activity $z_2$ per pair), while the exclusion rule
imposes that each edge of the lattice is shared by a total of
{\it at most} two particles. 
In Sect.\ 3.1, we present its expected phase diagram on a regular square or
honeycomb lattice obtained by applying the same ideas as above, i.e. relating the
different phases of the model to the compatibility between
the exclusion rules on one hand and the vertex-bicolorability of the lattice 
on the other. 
These ideas are tested in Sect.\ 3.2 where we solve the model on random
trivalent vertex-bicolorable lattices, by means of a six-matrix integral. 
The gravitational critical surface $g=g_c(z_1,z_2)$ is explicitly shown to contain the 
expected matter phase diagram, with in particular a critical Ising transition line
meeting a first order line at a tricritical Ising
transition point for some positive values of the activities 
$(z_1,z_2)=(z_1^{(t)},z_2^{(t)})$. It also displays a line of Lee-Yang edge
singularities terminating at a higher order critical point described 
by a non-unitary CFT
with central charge $c(2,7)=-68/7$  at some point $z_1=z_1^{(t')}>0$ and
$z_2=z_2^{(t')}<0$. 
A few concluding remarks are gathered in Sect.\ 4, while additional technical
derivations or more involved cases are left to appendices A-E.

\newsec{Nearest neighbor exclusion models on regular and random lattices} 

In this section, we study generic models of nearest neighbor exclusion
in which hard particles live on the vertices of either regular on random
lattices. The exclusion rule simply states that when a vertex is occupied
by a particle, all its nearest neighbors must be vacant. We first 
briefly recall known facts about the so-called ``hard square" model 
of hard particles on
the regular square lattice [\xref\BAX-\xref\GF] and the corresponding model
of hard particles on the regular honeycomb
lattice \RUN. As explained below, these models share the same qualitative phase 
diagram, where the structure of the ordered phase strongly relies on the
vertex-bicolorability (i.e. the bipartite nature) of the underlying
lattice. We then present an exact solution of the same models on
random lattices. We first consider the case of arbitrary random
graphs standardly used in the context of discretized 2D quantum gravity 
[\xref\DGZ,\xref\EY].
We note the disappearance of the ordered phase which can be traced back
to the generic lack of vertex bicolorability of the graphs. 
We therefore study the case of vertex-bicolorable (or so-called
Eulerian) random graphs, for which we derive the 
phase diagram
in the planar limit. As expected, the ordered phase is reinstated in
this case, and we find a continuous transition point in the universality
class of the critical Ising model coupled to 2D quantum gravity.

\subsec{Nearest neighbor exclusion on the square and honeycomb lattices}
\fig{Sample configurations of the hard particle model on the honeycomb (a)
and square (b) lattices. The exclusion constraint between particles 
(represented as black dots) may be translated into a non-overlapping
constraint for hard tiles, either hexagons (a) or squares (b).}{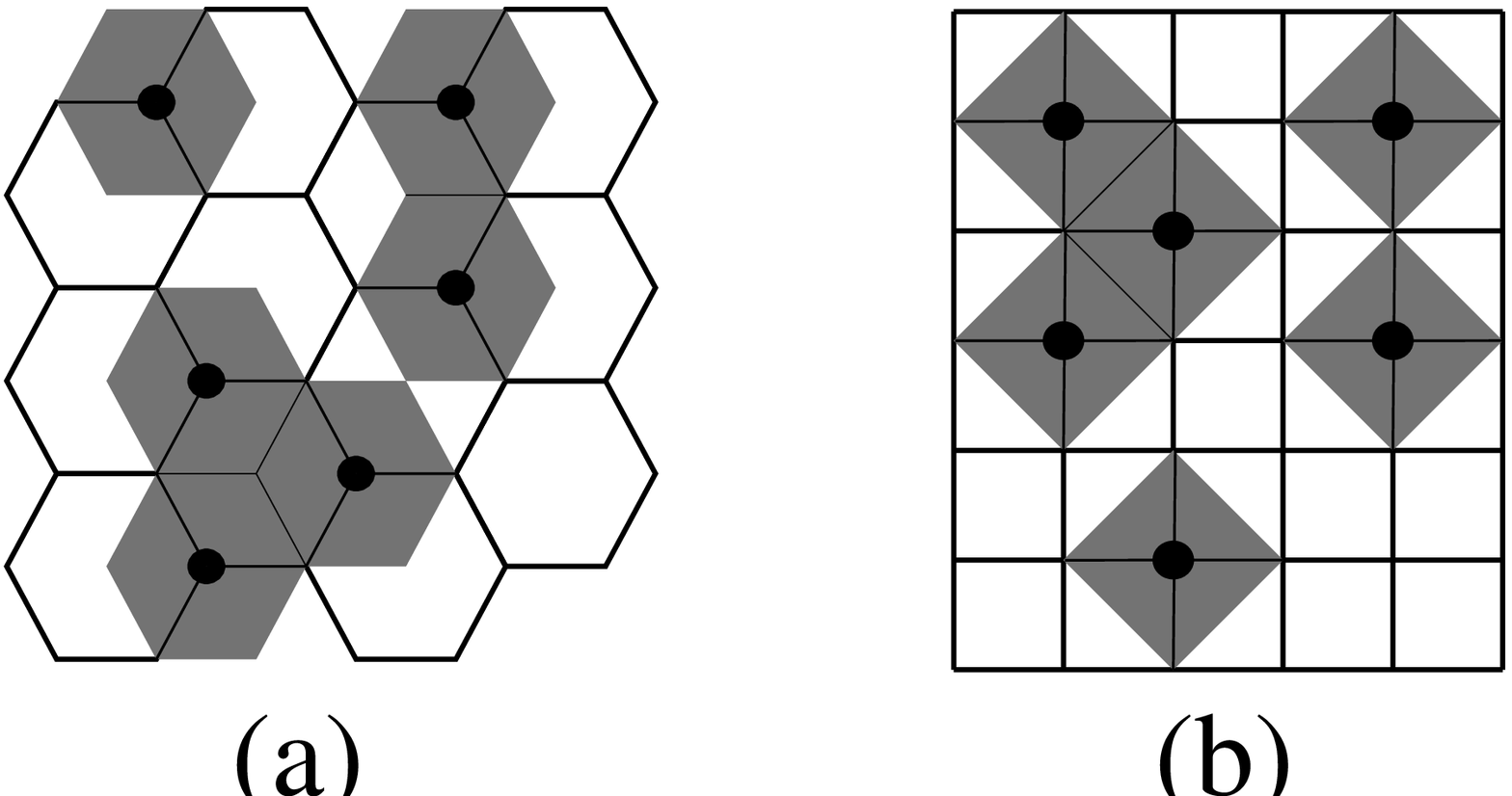}{10.truecm}
\figlabel\trisquare

Let us consider nearest neighbor particle exclusion models on 
respectively the regular trivalent honeycomb lattice and 
regular tetravalent square lattice. 
Each particle comes with an activity $z$ and excludes its
nearest neighbors. A simple pictorial representation of the exclusion
rule is to replace particles with non-overlapping hexagons, resp. squares, 
with centers on the vertices of the lattice, as depicted in Fig.\ \trisquare .
The phase diagrams of both models have been derived using numerical 
(corner) transfer matrix methods in [\xref\BAXHS,\xref\RUN]. 
They both display two critical points. 

The first ``non-physical" critical point occurs at a negative value 
$z_{-}$ of the activity and corresponds to the so-called Lee-Yang edge 
singularity \KF\ (non-unitary CFT with central charge 
$c(2,5)=-22/5$), 
with the estimates $z_{-}^{(3)}=-0.15\ldots$ (extracted from \BAX) 
for the (trivalent) honeycomb 
lattice and $z_{-}^{(4)}=-0.122\ldots$ \BAXHS\ for the (tetravalent) square lattice. 
The values $z_{-}$ of $z$ at these critical points can be obtained   
from the singularity of the thermodynamic free energy $F(z)$ expressed as an 
alternating series in powers of $z$. 
For $z \buildrel {\scriptscriptstyle >} \over {\to} z_{-}$, the singular part
of the free energy behaves as $F(z)\vert_{\rm sing}\sim (z-z_{-})^{2-\alpha}$ 
where $\alpha=7/6$ is the thermal
exponent predicted by the CFT. 
This apparently non-physical critical point
with negative activity can be reinterpreted as a positive activity critical
point at $t_{+}=-z_{-}$ for heaps of hexagons, resp. squares with activity $t=-z$ per 
object in $2+1$ dimensions, with free energy $\Phi(t)=-F(z=-t)$ \INV.

The second ``physical" critical point occurs at a positive value
$z_{+}$ and corresponds to a continuous transition between a low
activity disordered fluid phase and a high activity ordered crystalline 
phase where the hard particles condense preferably on one of the two
sublattices of the lattice. Estimates for the critical points are
$z_{+}^{(3)}=7.92\ldots$ \RUN\ and $z_{+}^{(4)}=3.7962\ldots$
\BAXHS.
More precisely, the honeycomb and square
lattice are bipartite lattices whose vertices can be naturally bicolored,
say black and white, in such a way that a vertex of one color has
only neighbors of the other. The corresponding order parameter 
$M=\rho_{B}-\rho_{W}$, which measures the difference of density
of particles between the black and white sublattice, is zero in the
fluid phase and non-zero in the crystalline one.
Based on numerical evidence, it is commonly accepted that this transition
lies in the universality class of the critical Ising model (CFT with
$c(3,4)=1/2$), with a thermal
exponent $\alpha=0$ and a magnetic exponent $\beta=1/8$ corresponding
to the singularity of the order parameter $M\sim (z-z_{+})^\beta$ for 
$z\buildrel{\scriptscriptstyle >}\over{\to} z_{+}$.

Note that the two above critical points are essentially different in nature.
The Lee-Yang critical point is much more universal and does not rely
on any particular geometrical feature of the lattice. It is also
observed for hard particles on the triangular lattice (the so-called
``hard-hexagon" model). On random lattices, it was first observed
for the Hard Dimer model in \HDIM\ and, as we shall see, it will show in all
the hard particle models studied throughout this paper.
On the contrary, the universality of the crystallization point strongly 
relies on the bicolorability of the underlying lattice. In the case
of the triangular lattice for instance, one finds instead a crystallization point
governed by the critical three-state Potts model (CFT with $c(5,6)=4/5$),
directly related to the tricolorability of the lattice.
In the case of random graphs, we shall also find that the bicolorability
is crucial to recover the Ising crystallization point.

\subsec{Nearest neighbor exclusion model on a random lattice}

We now turn to the study of nearest neighbor exclusion models on random graphs.
In this section we will concentrate on the case of hard particles living
at the vertices of 
arbitrary random {\it tetravalent} graphs. We use the standard matrix integral method
to generate all configurations of the exclusion model on all possible 
tetravalent fatgraphs, including a weight $g$ per tetravalent vertex
(empty or occupied), $z$ per particle, and the usual weight $N^{2-2h}$ for
graphs of genus $h$. We will mainly be interested in the planar limit
$N\to \infty$ that selects only graphs with the topology of the sphere $h=0$.
More precisely, the generating function reads 
\eqn\hsmod{\eqalign{ 
Z^{(4)}_N(g,z)&=\int dA dB e^{-N{\rm Tr}\, V(A,B)} \cr
V(A,B)&=-{1\over 2}A^2 +AB -g{B^4\over 4} -gz {A^4\over 4}\cr}}
where $A,B$ are Hermitian matrices with size $N\times N$, and the measure
is normalized so that $Z^{(4)}_N(0,0)=1$.
The Feynman diagrammatic expansion of \hsmod\ is readily seen to generate
tetravalent graphs with two types of vertices, occupied ($A^4/4$, with weight $gz$)
and empty ($B^4/4$ with weight $g$), and with the propagators given by the inverse
of the quadratic part of $V$, namely: 
\eqn\quadV{V_{\rm quad}={1\over 2}(A,B)\  Q \pmatrix{A\cr B} \quad 
{\rm and}\quad 
\pmatrix{\langle  AA\rangle&\langle  AB\rangle\cr\langle 
BA\rangle&\langle BB\rangle}
\propto Q^{-1}=\pmatrix{0 & 1 \cr 1 & 1}
}
hence
$\langle B_{ij}B_{kl} \rangle
=\delta_{il}\delta_{jk}/N$ and $\langle A_{ij}B_{kl} \rangle=\delta_{il}\delta_{jk}/N$
(for a general review on matrix models see [\xref\DGZ,\xref\EY] and references therein).
The vanishing of the $\langle AA\rangle$ propagator clearly enforces the exclusion rule.  

To compute \hsmod, we apply the standard technique of bi-orthogonal polynomials \DGZ.
We introduce the monic bi-orthogonal polynomials $p_n$, $q_m$ \wrt
the scalar product $(f,g)=\int dx dy e^{-NV(x,y)} f(x) g(y)$, satisfying
\eqn\monichs{ (p_n,q_m)=h_n \delta_{n,m} }
where the ``norms" $h_n$ are fully determined by the requirement 
that the polynomial  $p_n$ be monic of degree $n$. 
Introducing $h_n^{(0)}=h_n(g=0,z=0)$, and  
after reduction of \hsmod\ to an eigenvalue integral, we
may rewrite \DGZ\ 
\eqn\rewhs{ Z^{(4)}_N(g,z)=\prod_{n=0}^{N-1} {h_n \over h_n^{(0)}}}
which reduces the computation of the partition function to that of the $h_n$'s.

We introduce the operators of multiplication by eigenvalues
$Q_1$ and $Q_2$, expressed on the orthogonal polynomials
as $Q_1 p_n(x)=xp_n(x)$ and $Q_2 q_m(y) = y q_m(y)$, and the operators
of derivation \wrt eigenvalues $P_1$ and $P_2$ expressed as
$P_1 p_n(x)= p_n'(x)$, $P_2q_m(y)=q_m'(y)$. Integrating by parts, 
we get the system of equations
\eqn\sysths{ \eqalign{
(P_1p_n,q_m)&= N({\partial \over \partial x}V(x,y)p_n(x),q_m(y))=
N(p_n,Q_2q_m)-N((Q_1+gz Q_1^3)p_n,q_m) \cr
(p_n,P_2q_m)&= N({\partial \over \partial y}V(x,y)p_n(x),q_m(y))=
N(Q_1 p_n,q_m)-Ng(p_n,Q_2^3q_m) \cr}}
Note that $Q_1p_n$ is a linear combination of the $p_j$'s for $0\leq j \leq
n+1$, and similarly for $Q_2q_m$, combination of the $q_i$'s for $0\leq i\leq
m+1$, while $P_1p_n$ is a linear combination of the $p_j$'s for $0\leq j \leq
n-1$ and similarly $P_2 q_m$, combination of the $q_i$'s for $0\leq i\leq
m-1$. From orthogonality, we see that for $m>n+3$ in the first line of
\sysths\ and for $m<n-3$ in the second, we get respectively that
$(p_n,Q_2q_m)=0$ and $(Q_1 p_n,q_m)=0$. Hence the linear combinations reduce
respectively to finite ranges of indices $n-3\leq j\leq n+1$ and $m-3\leq i\leq m+1$.  
Moreover, from the parity of the potential $V(x,y)=V(-x,-y)$, we deduce
the parity of the polynomials: $p_n(x)=(-1)^n p_n(-x)$ and
$q_m(y)=(-1)^m q_m(-y)$. So finally the action of $Q_1$ and $Q_2$
takes the form
\eqn\actQ{\eqalign{ 
Q_1 p_n(x)&= p_{n+1}(x)+ r_n p_{n-1}(x)+s_np_{n-3}(x)\cr
Q_2 q_m(y)&= q_{m+1}(y)+ {\tilde r}_m q_{m-1}(y)+{\tilde s}_mq_{m-3}(y)\cr}}
The fact that $Q_1$ and $Q_2$ have a finite range is generic of multimatrix
models with polynomial interactions.
Eqns.\ \sysths\ and \actQ\ can be expressed in an operatorial way. Introducing 
the adjoint operators $Q_1^\dagger$ and $Q_2^\dagger$ \wrt the above scalar product,
eqn.\ \sysths\ takes the form:
\eqn\opsys{ \eqalign{
{P_1\over N}&= Q_2^\dagger -Q_1- gz Q_1^3 \cr
{P_2\over N}&= Q_1^\dagger -g Q_2^3 \cr}}
Let us introduce the shift operators $\sigma,\tau$ acting respectively on the
$p$'s and $q$'s  as $\sigma p_n=p_{n+1}$, $\tau q_n=q_{n+1}$ 
and their adjoints $\sigma^\dagger,\tau^\dagger$,
such that 
\eqn\defsigtau{
\sigma^\dagger=\tau^{-1} v\qquad \qquad \tau^\dagger =\sigma^{-1} v}
where $v=v^\dagger$ is the diagonal operator acting as $v p_n=v_n p_n$ and
$v q_n=v_n q_n$, with 
\eqn\defvh{v_n={h_n\over h_{n-1}}}
Analogously we define the diagonal operators $\nu,r,s,{\tilde r},{\tilde s}$
acting on $p_n$ and $q_n$ respectively as the multiplication by 
$n,r_n,s_n,{\tilde r}_n,{\tilde s}_n$. In terms of these operators,
the $P$ and $Q$ operators read finally
\eqn\opq{ \eqalign{
Q_1 &= \sigma +\sigma^{-1} r+\sigma^{-3} s\cr 
Q_2 &= \tau +\tau^{-1} {\tilde r}+\tau^{-3}{\tilde s}\cr
Q_1^\dagger &=\tau^{-1} v+ r v^{-1} \tau + s (v^{-1} \tau)^3\cr 
Q_2^\dagger &=\sigma^{-1} v+ {\tilde r}v^{-1}\sigma+ {\tilde s} (v^{-1}\sigma)^3\cr
P_1 &= \sigma^{-1} \nu + O(\sigma^{-3}) \cr
P_2 &= \tau^{-1} \nu+ O(\tau^{-3}) \cr}}
To obtain a system of recursion relations involving the sequence $v_n=h_n/h_{n-1}$,
let us now write order by order in $\tau$ and $\sigma$ the two relations
\opsys:
\eqn\durdur{\eqalign{
O(\sigma^{-1})&: \sigma^{-1}{\nu\over N}=\sigma^{-1}(v-r)-gz (r
\sigma^{-1}r+\sigma^{-1}r^2+(\sigma^{-1}r)^2\sigma+\sigma^{-1}s+\sigma^{-2}s\sigma+
\sigma^{-3}s\sigma^2)\cr 
O(\sigma)&:\ \ \  0= {\tilde r}v^{-1}\sigma -\sigma 
-gz(\sigma r+r \sigma+\sigma^{-1}r\sigma^2)\cr 
O(\sigma^3)&:\ \ \  0= {\tilde s} (v^{-1}\sigma)^3-gz \sigma^3\cr}}
and
\eqn\durmou{\eqalign{
O(\tau^{-1})&: \tau^{-1}{\nu\over N}=\tau^{-1}v -g ({\tilde r}\tau^{-1}{\tilde r}
+\tau^{-1}{\tilde r}^2+(\tau^{-1}{\tilde r})^2\tau+\tau^{-1}{\tilde
s}+\tau^{-2}{\tilde s}\tau+\tau^{-3}{\tilde s}\tau^{2})\cr
O(\tau)&:\ \ \  0= r v^{-1}\tau -g(\tau {\tilde r} 
+{\tilde r} \tau+\tau^{-1}{\tilde r}\tau^2)\cr
O(\tau^3)&:\ \ \ 0= s (v^{-1}\tau)^3-g \tau^3\cr}} 
It is not difficult to check that the first lines of  
\durdur\ and \durmou\ are equivalent modulo the other
equations, so that we are left with five equations for the five unknown 
sequences $v_n,r_n,s_n,{\tilde r}_n,{\tilde s}_n$.

The planar (genus zero) limit of these equations amounts to taking
$n,N\to \infty$ with $x=n/N$ fixed, in which case all sequences 
converge to functions of $x$. More precisely, we define the rescaled limits
$V(x),R(x),S(x),{\tilde R}(x),{\tilde
S}(x)$ of respectively $gv_n,gr_n,g^2s_n,g{\tilde r}_n,g^2{\tilde s}_n$. 
These functions are determined by
rewriting \durdur\ and \durmou\ in this limit which amounts to treating
all operators as scalars (in particular $\sigma=\tau=1$), with the result
\eqn\detersys{\eqalign{gx&=V-R-3z(R^2+S)\cr
0&= {\tilde R}-V(1+3 z R)\cr
0&= {\tilde S}-z V^3\cr
0&=R-3 V {\tilde R}\cr
0&=S-V^3\cr}}
After substitutions, this reduces to
\eqn\fineq{gx=\varphi(V)\equiv V(1-3zV^2)-{3 V^2 \over (1-9 z V^2)^2}}
For fixed $g$ and $z$ this equation defines upon inversion the function
$V(x)$ encoding the asymptotic properties of the sequence $v_n$, hence
those of the $h_n$'s. 
More precisely, using eqn.\ \rewhs,
the thermodynamic free energy in the planar limit reads
\eqn\thremof{\eqalign{ f_0^{(4)}(g,z)&=
-\lim_{N\to \infty} {\rm Log}(Z_N^{(4)}(g,z))/N^2 \cr
&=-\int_0^1 dx (1-x) {\rm Log}\left({V(x)\over g x}\right) \cr}} 
where the normalization $gx$ in the logarithm ensures the correct
normalization of the partition function, namely $f_0^{(4)}(g=0,z=0)=0$. 
For $g$ sufficiently small, eqn.\ \fineq\ determines a unique solution
$V(x)$ which is monotonous and such that $V(x)\sim gx$ for small $x$.
To compute $f_0^{(4)}(g,z)$, we must substitute this solution
into \thremof\ and perform the integration. It is natural to
perform the change of variables $x\to V$, which yields
\eqn\calcfre{f_0^{(4)}(g,z)=\int_0^{V_{g,z}}dV{\varphi'(V)\over g}
\left(1-{\varphi(V)\over g}\right)
{\rm Log}\left({\varphi(V)\over V}\right)} 
where $V_{g,z}$ is the value of $V$ at $x=1$, satisfying 
$\varphi(V_{g,z})=g$ for fixed $g$ and $z$.
The singularities of the planar free energy are due to those
of $V_{g,z}$ as a function of $g$ and $z$. 
For fixed $z$, the first singularity of $f_0^{(4)}(g,z)$ is attained at a critical
value $g=g_c(z)$, where the value $V_c(z)\equiv V_{g_c(z),z}$ 
is such that 
$\varphi'(V_c(z))=0$. In the vicinity of this point, \fineq\ reduces
to $g_c(z)-g \sim (V_c(z)-V_{g,z})^2$, which in turn yields a generic square root
singularity for $V_{g,z}$ in \thremof. To get the corresponding singularity of the free
energy, we note that,  
taking successive derivatives of \calcfre\ \wrt $g$ and using $\varphi(V_{g,z})=g$,
we simply get the general formula
\eqn\singfre{ {d^2\over dg^2} g^3 {d \over dg} f_0^{(4)}(g,z)=1-{g\over V_{g,z}}
{dV_{g,z}\over dg}}
The square root singularity of $V_{g,z}$ 
immediately translates into a string susceptibility exponent $\gamma$, defined by
$f_0^{(4)}(g,z)\vert_{\rm sing}\sim(g_c(z)-g)^{2-\gamma}$, with value $\gamma=-1/2$.
The critical value $g_c(z)$ also corresponds to the maximum value of $g$ for which 
the series expansion of the free energy converges.  
\fig{Critical line $g_c(z)$ for the hard particle model on arbitrary random
lattices, as obtained by setting $\varphi'(V)=0$. The solid line represents
the true critical values of $g=g_c(z)$ corresponding to the maximum
of $\varphi$ first attained by the change of variable $x\to V$, while
the dashed line corresponds to a further minimum as displayed on the top right
plot of the figure. These extrema merge at the Lee-Yang critical 
point $z_-$ where $\varphi''(V)=0$ (top center plot) and disappear from the real plane
(by becoming complex) for $z<z_-$ 
(top left plot).}{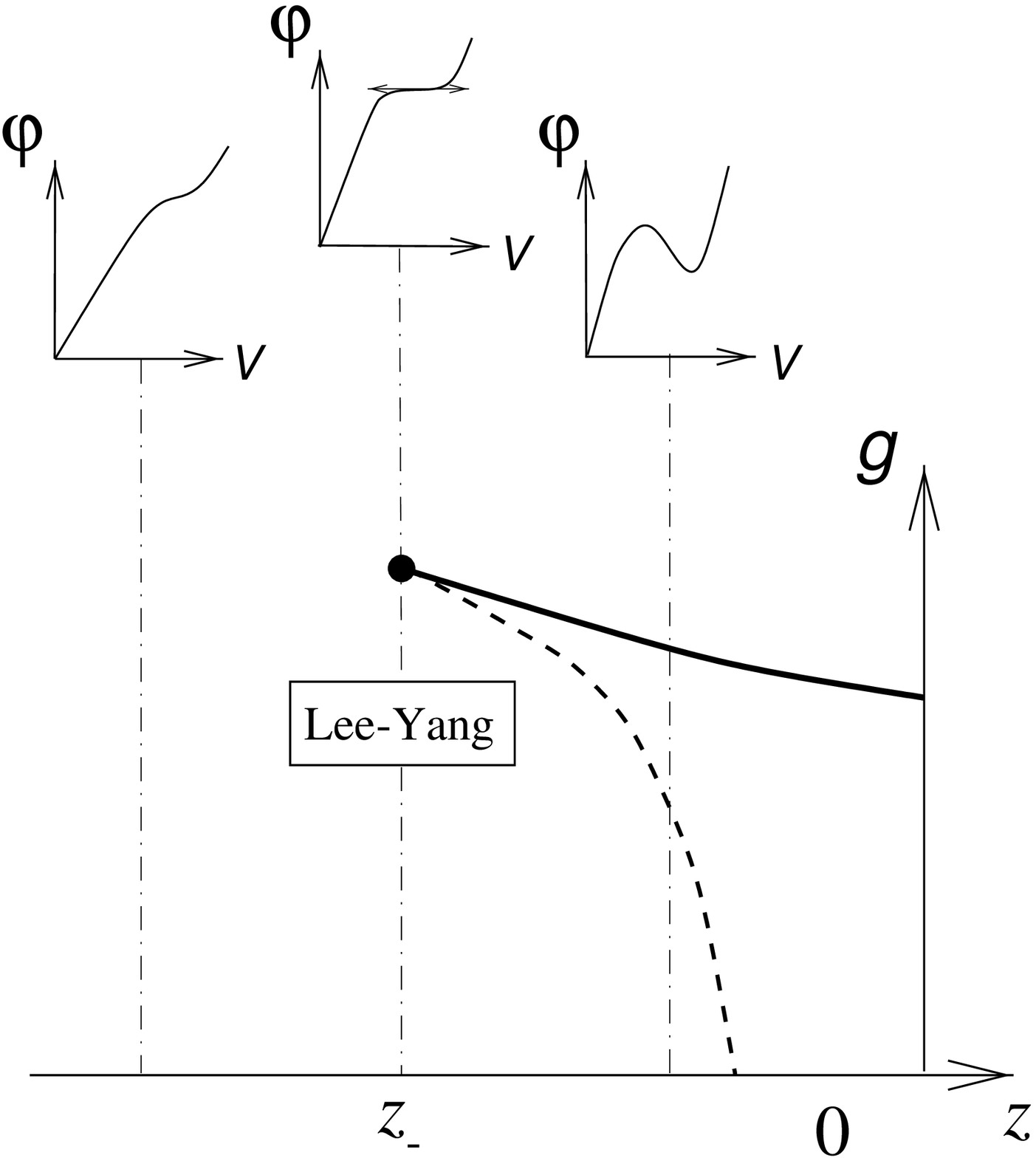}{9.truecm}
\figlabel\crilinb
Writing $g_c(z)=\varphi(V_c(z))$ and $0=\varphi'(V_c(z))$ yields parametric equations
for the critical line $g_c(z)$, in terms, say, of the parameter $u=3z V_c(z)^2$:
\eqn\parameq{ z={12 u(1+3u)^2\over (1-3 u)^8} \qquad g_c(z)={(1-3u)^4(1+10u-15u^2)\over
12 (1+3u)^2} }
This line is plotted in Fig.\crilinb. The solid curve corresponds to the true
critical values of $g=g_c(z)$, corresponding to a first maximum of $\varphi(V)$, while
the dashed line corresponds to a further minimum of $\varphi(V)$ which is never attained
by the above change of variables. The solid line stops at a finite negative value
of $z$ below which $g_c(z)$ becomes complex. 
We note that for $z=0$ ($u=0$) we recover the critical value $g_c=1/12$ for pure 
arbitrary tetravalent graphs \BIPZ.
The solid and dashed lines merge into a cusp at a higher order 
critical point satisfying in addition
$\varphi''(V)=0$, as it corresponds to the coalescence of the
maximum and the minimum of $\varphi$ (see Fig.\crilinb). 
This point corresponds to the values
\eqn\twosol{\eqalign{
z_-&=-{25\over 8192}(11\sqrt{5}+25)=-0.151... \cr
g_-&=g_c(z_-)={64\over 45}(13\sqrt{5}-29)=0.0979... \cr
V_-&=V_c(z_-)={32/75}(7\sqrt{5}-15)=0.278...\cr}}
At this point, the above scaling argument now becomes
$g_--g\sim (V_--V)^3$, hence translates directly into a string susceptibility
exponent $\gamma=-1/3$.
As the critical activity $z_-$ is negative, we identify this higher
critical point with the unique non-unitary theory with $\gamma=-1/3$, namely
the Lee-Yang edge singularity with $c(2,5)=-22/5$, coupled to 2D quantum gravity
(see Appendix A).
This identification may be corroborated in two different ways. On one hand, one may
compute the thermal exponent $\alpha$ which measures the singularity of the free energy
at the critical point as a function of $z$, and compare it to the predicted value
$\alpha=1/2$ for the Lee-Yang edge singularity coupled to 2D quantum
gravity, as re-derived in Appendix A below. 
Writing the thermodynamic free energy per site as 
$f^{(4)}(z)=-\lim_{A\to\infty}{1\over A} {\rm Log}\, Z_A(z)$
where $Z_A(z)$ is the partition function for planar graphs of fixed number 
of vertices (area) $A$,
we read $\alpha$ from the singular part $f^{(4)}(z)\vert_{\rm
sing}\sim(z-z_-)^{2-\alpha}$. Using $f^{(4)}(z)={\rm Log}\, g_c(z)$, $\alpha$
may be obtained by computing the singular part of $g_c(z)$. 
Expanding eqn.\ \parameq\ in powers
of $u-u_-$, we find that $g_c(z)-g_-= a (u-u_-)^2 +b (u-u_-)^3+...$ and
$z-z_-=a' (u-u_-)^2 +b' (u-u_-)^3+...$ with $u_-=3z_- V_-^2$, and which, upon inversion,
yields $g_c(z)-g_-=a''(z-z_-)+b''(z-z_-)^{3\over 2}+...$ with a non-vanishing value of
$b''$. Hence we get $2-\alpha=3/2$ as expected.
On the other hand, one may also derive the so-called double 
scaling limit of the model at the
critical point and write a differential equation for the renormalized string
susceptibility, easily identified with that of the Lee-Yang edge 
singularity coupled to 2D quantum gravity.
This derivation is presented in Appendix B below. 

Note that for $z<z_-$, the gravitational critical value of $g$
becomes complex with $g_c(z)=\rho e^{i\theta}$, but it still governs 
the large area behavior of the above partition function $Z_A(z)$ which
now oscillates typically as 
\eqn\oscillo{ Z_A(z)\sim \rho^{-A} \cos(A\theta)}
which allows to identify the thermodynamic free energy as ${\rm Log}\,\rho$.  

In conclusion, when comparing with the regular lattice results, 
we see that the naive gravitational version of the exclusion model
fails to reproduce the crystallization transition, and leaves us only
with the ``non-physical" Lee-Yang edge singularity. As shown in Appendix C,
the case of trivalent graphs instead of tetravalent is exactly solvable as well
and displays the same structure.  
The absence of a crystalline ordered phase
should not come as a surprise, as the partition function involves a sum over
graphs that are not generically bicolorable, hence do not allow for a canonical
crystalline order, where half of the vertices are preferentially occupied. 
To emphasize the role played by bicolorability for exclusion models, we note
that for large $z$ ($u \to 1/3$ in eqn.\ \parameq) the quantity 
$\sqrt{z}\,  g_c(z)$  tends to $2/9$, which is precisely the critical value
of $g$ in a model of pure bicolorable tetravalent graphs\foot{This critical value
was computed in \CK\ in the context of a particular O(n=1) gravitational model.}. 
This clearly shows
that in this limit the selected configurations
are half-occupied vertex-bicolorable graphs.    

To recover a crystallization transition at finite $z$, we shall 
consider in the next section the coupling of exclusion models to Eulerian gravity.

\subsec{Nearest neighbor exclusion models on vertex-bicolorable
random lattices}

In this section, we consider a restricted gravitational version of the nearest neighbor
exclusion model, in which we explicitly sum only over the so-called random 
Eulerian graphs, simply defined as vertex-bicolorable (or bipartite) graphs.
It turns out that the case of trivalent Eulerian graphs is technically simpler than that
of tetravalent ones, yet it displays the same qualitative physical behavior.
Therefore we will now concentrate on the trivalent case and leave 
the tetravalent one to Appendix E.

The configurations of the nearest neighbor exclusion model on
Eulerian trivalent graphs are again generated by a matrix model replacing 
\hsmod.
We now need a total of four matrices, as the
vertices must be bicolored and empty or occupied.
More precisely, we use a matrix $A_1$ (resp. $A_4$) for empty black
(resp. white) vertices and a matrix $A_2$ (resp. $A_3$) for occupied
white (resp. black) vertices. The resulting matrix model reads
\eqn\htmodfour{\eqalign{
Z^{(3)}_N(g,z)&=\int dA_1 dA_2 dA_3 dA_4 e^{-N{\rm Tr}\, V(A_1,A_2,A_3,A_4)} \cr
V(A_1,A_2,A_3,A_4)&=A_1 A_2 -A_2A_3 +A_3 A_4 -g({A_1^3\over 3}+{A_4^3\over 3})
 -gz({A_2^3\over 3}+{A_3^3\over 3})\cr}}
The quadratic form in $V(A_1,A_2,A_3,A_4)$ has been engineered so as to reproduce
the correct propagators, namely that only black and white vertices are connected
in the Feynman diagrams ($\langle A_i A_j\rangle =0$ if $i$ and $j$ have the same parity)
and that two occupied neighboring vertices exclude one-another
($\langle A_2 A_3\rangle=0$).

Due to the chain-like interaction between the matrices, this model turns out to 
be solvable by means of bi-orthogonal polynomials. In addition the symmetry
$A_{i}\leftrightarrow A_{5-i}$ implies that the two sets of polynomials are 
identical. We therefore introduce the
monic polynomials $p_n$, orthogonal \wrt
the appropriate symmetric scalar product, namely
\eqn\orthopol{ (p_n,p_m)=\int dx_1 dx_2 dx_3 dx_4
e^{-NV(x_1,x_2,x_3,x_4)} p_n(x_1) p_m(x_4)=  h_n \delta_{n,m} }
As before, we introduce operators $Q_i$ of multiplication by $x_i$, $i=1,2,3,4$
but this time all acting on $p_n(x_1)$, but the symmetry $A_{i}\leftrightarrow A_{5-i}$
immediately implies that $Q_3=Q_2^\dagger$ and $Q_4=Q_1^\dagger$. We also
introduce the operator $P_1$ acting on $p_n(x_1)$ as $d/dx_1$.  
These operators satisfy the system
\eqn\derto{ \eqalign{
{P_1\over N}&= Q_2 -g Q_1^2 \cr
Q_1&= Q_2^\dagger +g z Q_2^2 \cr}}
obtained by integrating by parts.

To write explicitly the action of $Q_1$ and $Q_2$ on the $p_n$'s, let us 
first notice that the potential $V$ satisfies the symmetry relation
\eqn\simV{ V(\omega x_1, \omega^2 x_2, \omega x_3,\omega^2 x_4)=V(x_1,x_2,x_3,x_4),
\qquad \omega=e^{2i\pi/3} }
This translates into the symmetry relation for the monic orthogonal polynomials
\eqn\sympn{ \omega^{2n} p_n(\omega x)= p_n(x) } 
Now from \derto\ it is easy to show that $Q_1$ and $Q_2$ have finite range, and more
precisely, thanks to the symmetry relation \sympn\
\eqn\ragepq{\eqalign{
Q_1&= \sigma+ \sigma^{-2}r^{(1)}+\sigma^{-5}r^{(2)}+\sigma^{-8}r^{(3)} \cr
Q_2&=  \sigma^2 s^{(0)}+\sigma^{-1} s^{(1)}+\sigma^{-4}s^{(2)} \cr}}
where $\sigma$ is the shift operator acting on the $p$'s as $\sigma p_n=p_{n+1}$
and the operators $r^{(i)},s^{(i)}$ are diagonal.
Introducing as before the diagonal operator $v$ with entries 
$v_n =h_n/h_{n-1}$, we have $\sigma^\dagger=\sigma^{-1} v$,
and therefore  
\eqn\adjqto{ Q_2^\dagger =s^{(0)} (\sigma^{-1} v)^2 +
s^{(1)} v^{-1} \sigma+ s^{(2)} (v^{-1} \sigma)^4 }
Expanding the relations \derto\ order by order in $\sigma$, we finally arrive at
\eqn\internat{\eqalign{
s^{(0)}&= g \cr
s^{(2)}&= -g^3 z \sigma^4 (\sigma^{-1} v)^4\cr
v&= s^{(1)} +g^2 z v( \sigma s^{(1)} \sigma^{-1} +\sigma^{-1}s^{(1)}\sigma)\cr
r^{(1)}&= g \sigma v\sigma^{-1} v+ gz \sigma s^{(1)} \sigma^{-1}s^{(1)}+g^2 z(
s^{(2)}+\sigma^{-2} s^{(2)} \sigma^2 ) \cr
{\nu\over N}&= s^{(1)} -g(r^{(1)}+\sigma^{-1}r^{(1)}\sigma)\cr }}
where $\nu$ is the diagonal operator with entries $n$.

In the planar limit, we express the system \internat\
in terms of the rescaled limiting functions
$S(x)= \lim g^2 z s^{(1)}$, $V(x)=\lim g^2 z v$, 
where $x=n/N$ while $n,N\to\infty$, and also use the fact that $\sigma\to 1$.
The third line of \internat\
allows to solve for $S=V/(1+2 V)$, so that we finally get
\eqn\finmas{ g^2z^2 x\equiv \varphi(V)= z {V\over (1+2V)^2} -2 V^2(1-2 V^2) }
Writing 
\eqn\criline{ \varphi'(V)= {(1-2V)(z-4V(1+2V)^4)\over (1+2V)^3} =0 }
we get two candidates for the critical line, namely 
$V=1/2$ or $z=4V(1+2V)^4$.
\fig{Critical line $g_c(z)$ for the hard particle model on vertex bicolorable 
random lattices, as obtained by setting $\varphi'(V)=0$. The solid line 
represents the true critical values of $g=g_c(z)$ corresponding to the maximum
of $\varphi$ first attained by the change of variable $x\to V$, while
the dashed line corresponds to a further minimum as displayed on the top
plots. The curve terminates at a Lee-Yang critical point similar to that
of Fig.\ \crilinb. The maximum and minimum swap determinations
at the critical Ising point $z_+$ where $\varphi''(V)=0$ 
(top center plot).}{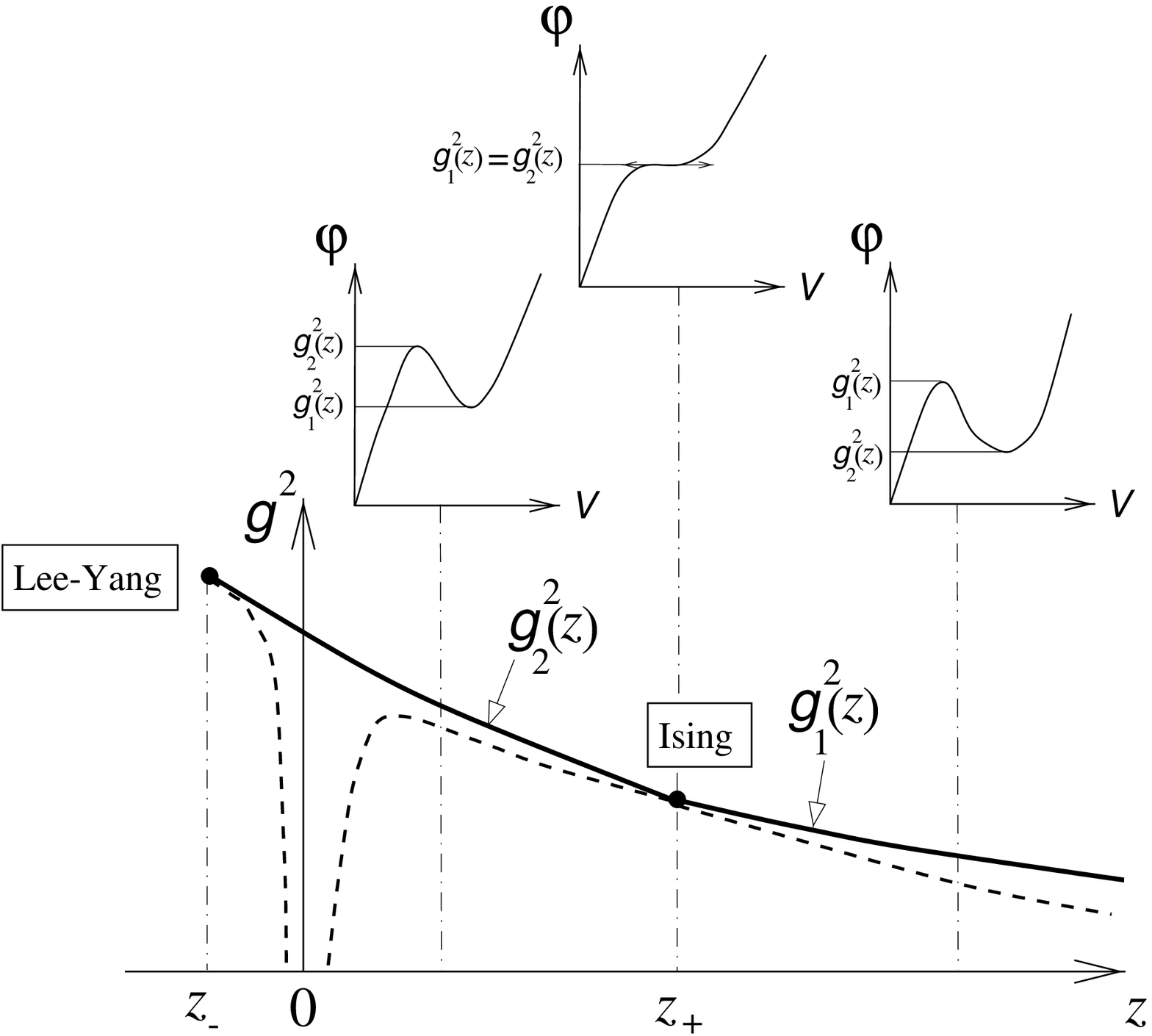}{11.cm}
\figlabel\linetri
\noindent These lead respectively to the two possible critical curves 
$g^2(z)=\varphi(V)/z^2$ parametrized by 
\eqn\twocril{\eqalign{ g_1^2(z)&={1\over 8 z}-{1\over 4 z^2}, 
\ \ z>0, \ \ V=1/2\cr
g_2^2(z)&={1+8V+10 V^2\over 8(1+2V)^8}, \ {\rm with} \ z(V)=4V(1+2V)^4 \cr}}
where the condition $z>0$ in the first line simply follows from the positivity
requirement for the norm ratios $v_n$, implying that $V$ and $z$ have the same sign. 
\fig{Critical lines in the $(V,z)$ plane as obtained by setting $\varphi'(V)=0$.
The correct line corresponds to the lowest value of $\vert V\vert$ and is
represented by a solid line. The Lee-Yang critical point $z_-$ is 
characterized by $dz/dV=0$ and corresponds to the merging and annihilation
of two extrema. The Ising critical point $z_+$ corresponds to the
crossing of two determinations of $V(z)$.}{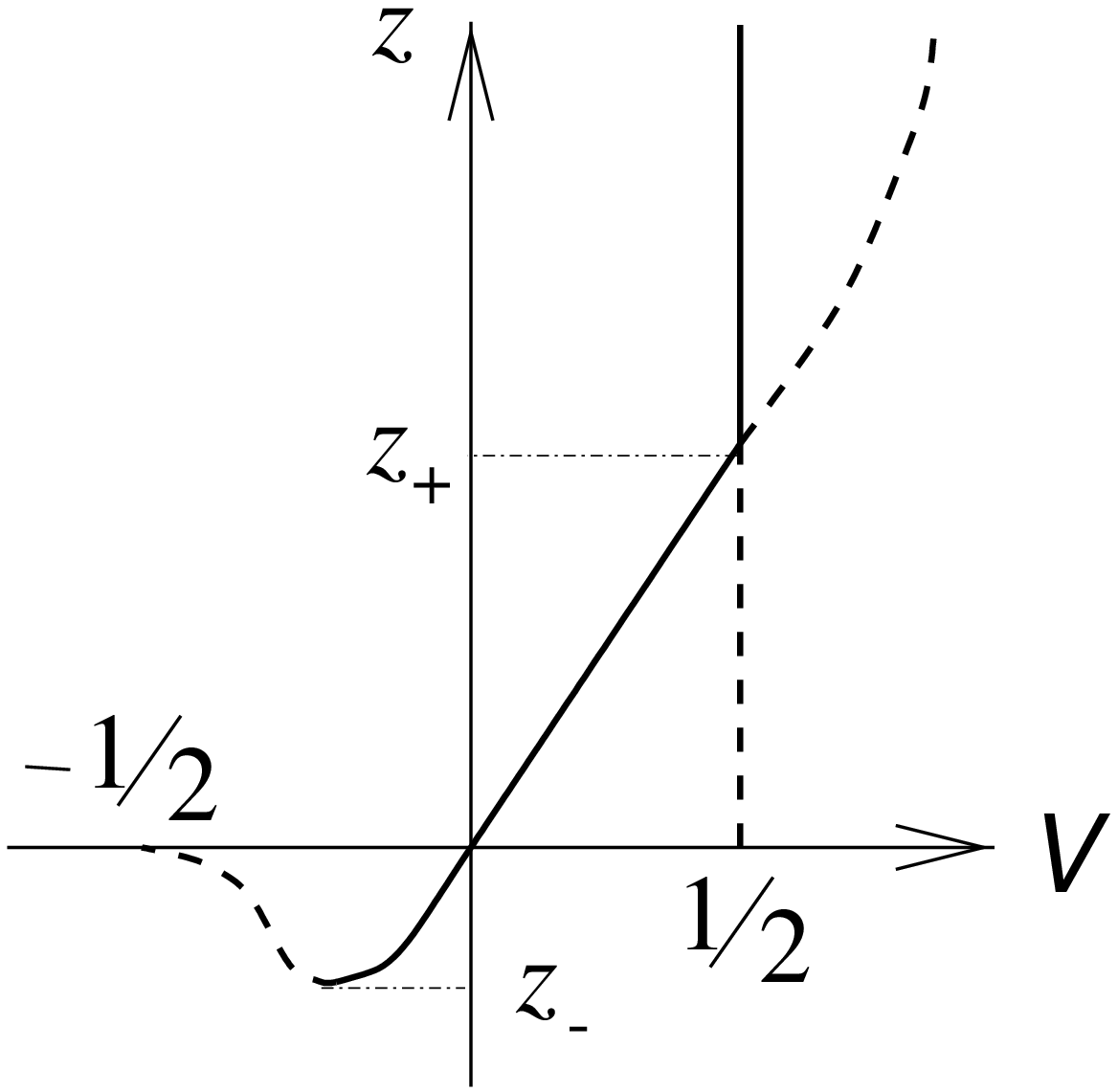}{6.truecm}
\figlabel\Vzdiag
These two lines are represented in Fig.\linetri. 
The choice of the correct determination in \twocril\ is best 
seen by plotting the critical lines in a $(V,z)$ diagram as in Fig.\ \Vzdiag.  
In this picture, the correct singularity of the free energy at fixed $z$
is always given by the lowest critical value of $\vert V\vert$ as 
it is the one attained by the above change of variables. 
The true critical points of the free energy are represented by solid pieces 
of curves in Fig.\ \linetri, and correspond to maxima of $\varphi(V)$, while the 
dashed parts correspond to further minima of $\varphi(V)$ never attained by 
the change of variables as before. 
Along these lines the string susceptibility exponent
is $\gamma=-1/2$. 
We note that for $z=0$, we recover the critical value $g_c^2=1/8$ for 
pure bicolored trivalent graphs [\xref\TUT,\xref\DEG]. 
Also for $z\to \infty$ we get 
$g^2(z)=g_1^2(z)\sim 1/(8z)$ as expected for half-occupied trivalent
bicolored graphs. 

The higher order critical points
correspond to the cuspidal singularity at $z_-$, where the maximum and minimum
of $\varphi(V)$ merge and annihilate each  other, and to the crossing between the 
two curves at $z_+$, where the value of the maximum of $\varphi$ hops from
$g_2(z)$ to $g_1(z)$ while its minimum hops from $g_1(z)$ to $g_2(z)$. 
The position of these points is obtained by writing the extra
condition $\varphi''(V)=0$, with 
\eqn\secder{\varphi''(V)= \left\{ \matrix{ {32-z\over 4} & {\rm if} & V=1/2\cr
4(2V-1)(1+10V) & {\rm if} & z=4V(1+2V)^4 \cr} \right. }
We get the two critical values 
\eqn\tripoint{ \eqalign{ (1) \ \ V_+&={1\over 2}, \qquad z_+=32, 
\qquad g_+^2={15 \over 2^{12}}\cr
(2) \ \ V_-&=-{1\over 10}, \qquad z_-=-{2^9\over 5^5},\qquad 
g_-^2={3.5^7\over 2^{20}}\cr}}
The critical point $z_-$ is similar to that of Sect.\ 2.2 and corresponds to
the Lee-Yang edge singularity.
The crucial outcome of our calculation is the emergence of a crystalline phase
at finite values of $z$, with a crystallization transition at $z_+$.
This critical point turns out to be in the universality class
of the critical Ising model on random graphs.
The simplest reason for these identifications
is that both above critical points have string susceptibility exponent
$\gamma=-1/3$ by construction, and that only two CFT's are candidates to
describe this\foot{Indeed for a CFT with central charge $c(p,q)=1-6(p-q)^2/(pq)$, we have
$\gamma=-2/(p+q-1)$, hence here $p+q=7$, and $(p,q)=(2,5),(3,4)$ only.}, 
the Lee-Yang edge singularity with $c(2,5)=-22/5$ which is non-unitary
as expected for a negative value $z=z_-$, and
the critical Ising model with $c(3,4)=1/2$ which is a unitary theory, as 
expected for a positive critical activity $z_+$.
To further confirm the identification of the critical Ising universality class,
one can compute the thermal exponent $\alpha$ for the 
crystallization transition. It is easy to see from the above formulas that
the transition from the curve $g_1(z)$ to $g_2(z)$ is continuous at $z_+$, with continuous
first and second derivative, and with a discontinuity of the third one. 
This gives a thermal exponent $\alpha=-1$, as expected for the critical Ising
model coupled 2D quantum gravity (see Appendix A). 
We have also derived the differential equations for the string susceptibility
for both cases in the corresponding double scaling limits, and identified them 
with the known results for the Lee-Yang and Ising critical models coupled to
2D quantum gravity. The details of these calculations are given in 
Appendix D below.

\newsec{Two-particle exclusion models on regular and random lattices}

In this section, we extend our analysis of exclusion models on bicolorable lattices
to incorporate the physics of higher order critical points. More precisely, 
in the previous section we have reproduced the universality class of the critical
Ising model within the context of exclusion models on bicolorable random graphs. 
The crucial feature leading to the Ising symmetry is the existence 
of two degenerate symmetric crystalline groundstates playing the role of the two
(up and down) ferromagnetic groundstates. 

We now wish to construct in the language of
particle exclusion a model reproducing the physics of the tricritical Ising model.
In the framework of spin systems,
the latter is found for instance in the phase diagram of a 
dilute Ising model, with spins $\sigma=0,\pm 1$ \BEG, 
where the spins $\sigma=0$ play the role
of annealed non-magnetic vacancies. The existence of a tricritical point is 
associated to that of three non-symmetric groundstates of constant spin:
the two groundstates $\sigma=+1$ and $\sigma=-1$ play symmetric roles, but $\sigma=0$
is on a different footing. The Ising second order phase transition in the absence of
vacancies extends into a line of second order transition points for low enough activity
per vacancy. On the other hand, at low enough temperature, a first order transition line 
separates a ferromagnetic phase at low activity per vacancy from a paramagnetic
phase at large activity per vacancy. Both lines merge at a tricritical point whose
behavior defines the tricritical Ising universality class (CFT with $c(4,5)=7/10$).
(For a general review on tricritical points in the context of spin systems or
lattice gases see \DG.)

In the next section, we show how to realize
a similar behavior within the framework of hard objects, by use of a
{\it two-particle} nearest neighbor exclusion model. We first describe its 
expected phase diagram on a regular bicolorable lattice. Here again, the bicolorability of
the underlying lattice is crucial for the existence of an ordered crystalline phase.
We then derive an exact solution of the same model on bicolorable random lattices
and recover the expected phase diagram, with in particular a tricritical point
coupled to 2D quantum gravity.

\subsec{Two-particle nearest neighbor exclusion models on the square and honeycomb
lattices}  

\fig{Sample configurations of the two-particle exclusion model on the 
honeycomb (a) and square (b) lattices. The black dots represent singly
occupied vertices while the circled black dots represent doubly occupied
vertices. The two-particle exclusion constraint that an edge be shared
by at most two particles may be translated into a non-overlapping
constraint for hard tiles, either hexagons for double occupancy and triangles 
for single occupancy (a) or big tilted squares for double occupancy
and small squares for single occupancy (b).}{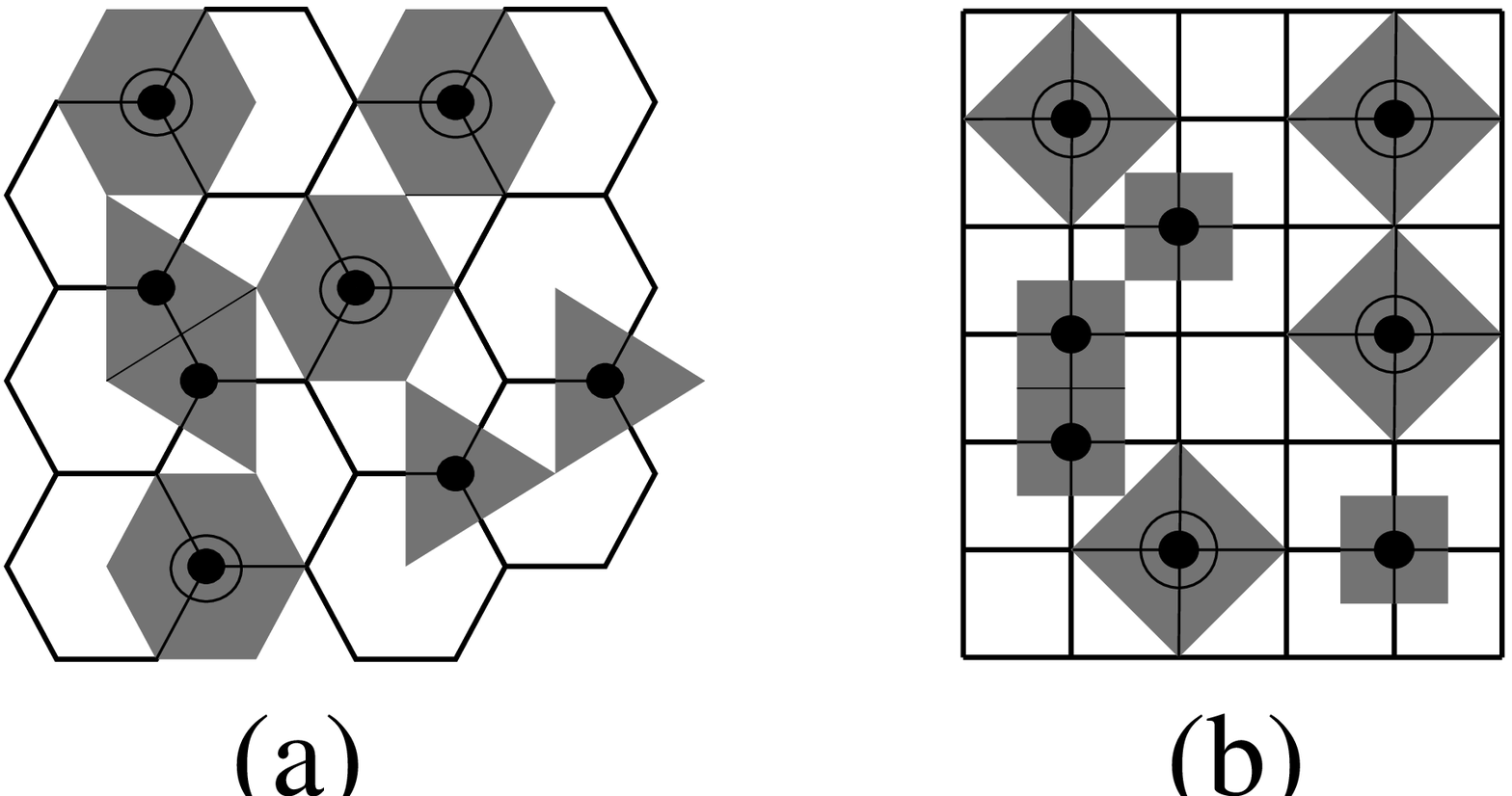}{10.cm}
\figlabel\twopart

In its simplest formulation, the two-particle exclusion model we wish to study is
defined by putting particles on the vertices of a regular bicolorable
lattice (typically square or honeycomb), with the exclusion rule that a total of 
at most two particles may occupy the two vertices adjacent to any edge. In particular,
we allow for two particles to occupy the same vertex, in which case its nearest neighbors
must be vacant, while singly occupied vertices may be nearest neighbors. We assign a weight
$z_1$ for singly occupied vertices and a weight $z_2$ for doubly occupied ones, with
typically $u=z_2/z_1^2$ measuring the attractive ($u>1$) or repulsive ($u<1$)
interaction between particles on the same vertex.
As in the one-particle models of Sect.\ 2, we may represent pictorially the exclusion
constraint in terms of non-overlapping hard objects on the corresponding lattices.
As shown in Fig.\ \twopart, for the square lattice, doubly occupied sites may be
represented as usual hard squares, which exclude their four neighbors, while
singly occupied sites are represented by a new type of squares which are twice as small
and may thus occupy two neighboring sites.
Similarly, for the honeycomb lattice, we represent doubly and  singly occupied
vertices respectively by hard hexagons and triangles that are twice as small.

\fig{Expected phase diagram in the $(1/z_2,u^{-1}=z_1^2/z_2)$ plane for 
positive activities. The ordered phase $M\neq 0$ corresponds
to a crystallization of the doubly occupied vertices on one of the two
mutually excluding sub-lattices.
This phase is separated from the fluid
phase $M=0$ by a second order critical line for small enough $u^{-1}$ 
and by a first order transition line for small enough $1/z_2$. Both lines
are expected to meet at a tricritical point. The three natural groundstates in
the problem are recovered along the axis $1/z_2=0$: for $u^{-1}<1$ the two
maximally filled doubly occupied configurations dominate (tilings with the
bigger tiles) while at $u^{-1}>1$, the full occupation by single
particles (tiling with smaller tiles) dominates. The latter groundstate
degenerates into a disordered fluid phase as soon as $1/z_2>0$.}{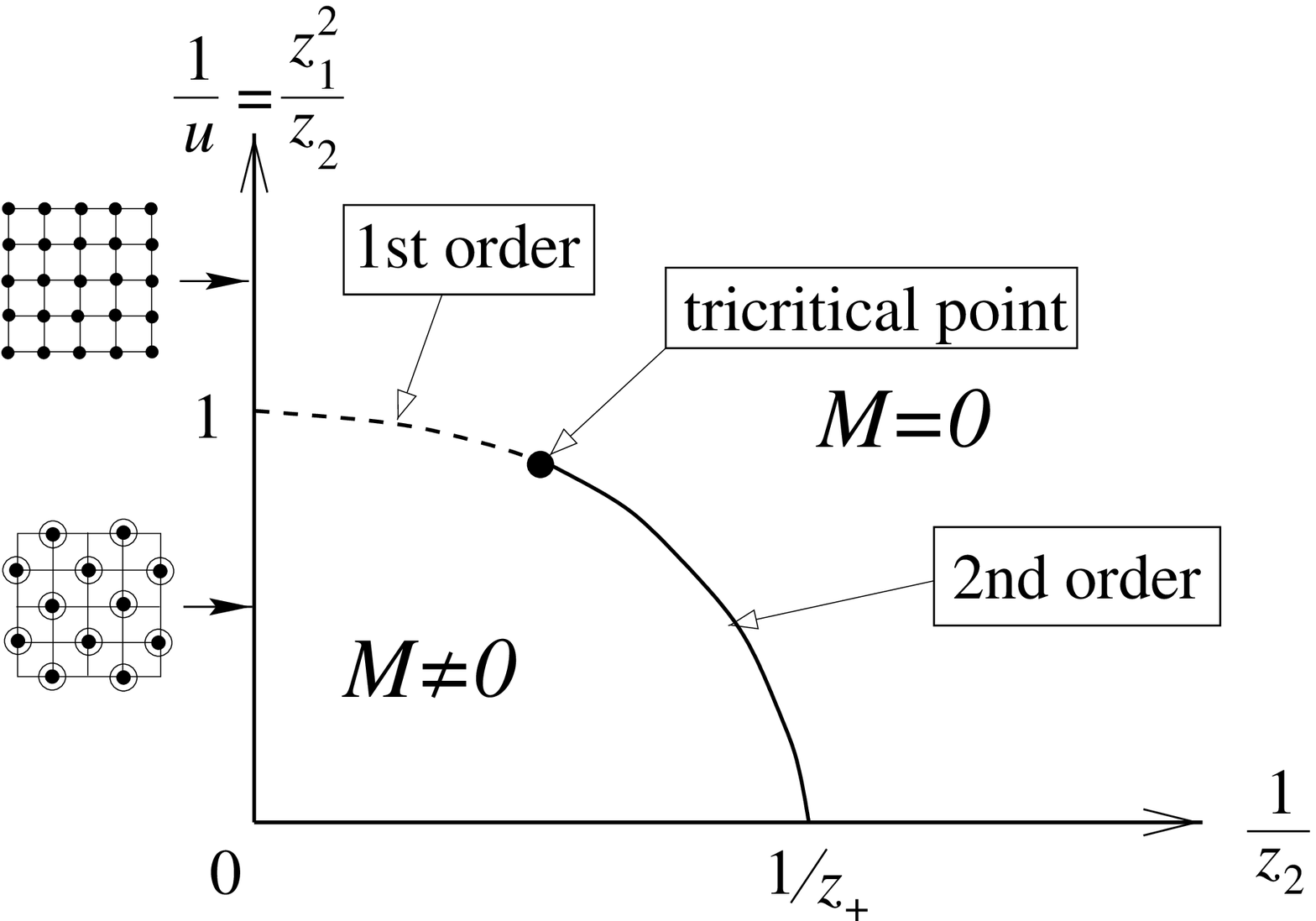}{13.truecm}
\figlabel\phasediag
In both cases, we have three crystalline groundstates corresponding to a maximal
covering of the lattice: two of them are symmetric and
use only the larger tiles which occupy one
of the two sublattices of the bipartite lattice,
the third one is obtained by tiling the lattice with the smaller objects.
The phase diagram of the model is best represented in the variables $(1/z_2,
u^{-1}=z_1^2/z_2)$ (see Fig.\ \phasediag). 
On the axis $u^{-1}=0$ (i.e. $z_1=0$), we recover 
the one-particle exclusion models of Sect.\ 2  with $z=z_2$, and in 
particular a
crystallization point at $z_2=z_+$. For fixed small enough $u^{-1}$, we expect
a similar second order transition at some $z_2=z_+(u)$. The critical curve $z_2=z_+(u)$
separates the liquid phase from the crystal phase, the latter
being characterized by the non-vanishing
of the order parameter $M=\rho^{(2)}_B -\rho^{(2)}_W$ expressing the difference
of densities of doubly occupied sites of either color. 
On the other hand, on the axis $(1/z_2)=0$, i.e. $z_2\to\infty$ and $z_1\to\infty$,
with $u^{-1}=z_1^2/z_2$ fixed, we have a competition between the three groundstates
of maximal occupation. The two symmetric groundstates made of larger tiles
have free energy per site $({\rm Log}\, z_2)/2$, while the other groundstate
made of smaller tiles has free energy ${\rm Log}\, z_1$. This leads to a first order
transition at $u=1$, with same order parameter $M=\pm 1$ for $u>1$ and $M=0$ for
$u<1$. We expect this transition point to extend into a first order transition curve
for small enough $(1/z_2)$. By analogy with the tricritical Ising model phase diagram,
we expect the two curves to meet at a tricritical point with $c(4,5)=7/10$.

These models do not seem to be simply solvable by integrable techniques, and the above
phase diagram is somewhat conjectural. However, in the following section we shall present
an exact solution of the random bicolorable lattice version of the two-particle
exclusion model, and show that it displays precisely the physical picture
described above.

\subsec{Two-particle exclusion model on vertex-bicolorable random lattices}

\fig{The non-vanishing propagators corresponding to
the six-matrix model generating the configurations of 
the two-particle exclusion model on vertex bicolorable random lattices.
Identifying respectively $A_1$, $A_3$, $A_5$ with empty,
singly occupied and doubly occupied black vertices on one hand
and $A_6$, $A_4$, $A_2$ with empty,
singly occupied and doubly occupied white vertices, the selected propagators
enforce both the bicoloring constraint (black vertices are connected to white 
vertices only and conversally) and the exclusion constraint (at most two
particles may share the same edge).}{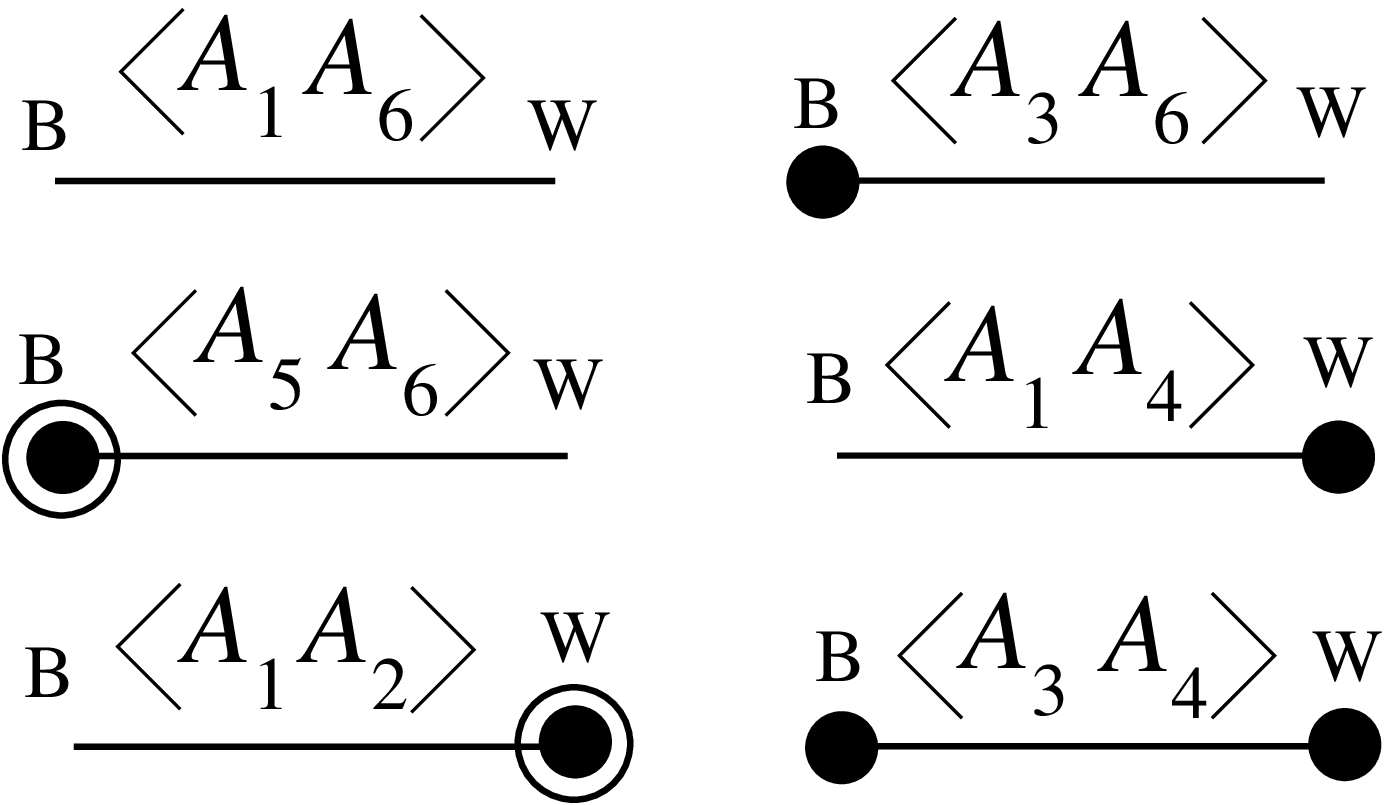}{8.cm}
\figlabel\propagsix 
We start with the matrix model
\eqn\dnneint{\eqalign{
Z_N(g,z_1,z_2)&=\int \prod_{i=1}^6 dA_i e^{-N{\rm Tr}\, V(A_1,A_2,A_3,A_4,A_5,A_6)}
\cr
V(A_1,A_2,A_3,A_4,A_5,A_6)&=\, A_1A_2-A_2A_3+A_3A_4-A_4A_5+A_5A_6 \cr
&-\, g({A_1^3\over 3}+{A_6^3\over 3})-gz_2 ({A_2^3\over 3}+{A_5^3\over 3})-gz_1
({A_3^3\over 3}+{A_4^3\over 3})
\cr}}
where as for Sect.\ 2.3, the $N\times N$ 
Hermitian matrices $A_i$ with odd (resp. even) index
correspond to black (resp. white) vertices,  
which can be empty ($A_1,A_6$), singly occupied ($A_3,A_4$) or doubly
occupied ($A_5,A_2$). It is easy to check that the inverse of the quadratic form
in $V$ generates the expected non-vanishing propagators $\langle A_1 A_2\rangle$,
$\langle A_1 A_4\rangle$, $\langle A_1 A_6\rangle$, $\langle A_5 A_6\rangle$,
$\langle A_3 A_6\rangle$, $\langle A_3 A_4\rangle$ (see Fig.\propagsix).
Remarkably enough, the matrix interaction in $V$ is simply chain-like, allowing for a
solution using bi-orthogonal polynomials.

As before, the symmetry $A_i\leftrightarrow A_{7-i}$ of
$V$ ($i=1,2,3$) ensures that the left and right polynomials are identical,
hence we define the monic orthogonal polynomials $p_n$ \wrt the appropriate 
symmetric scalar product, namely
\eqn\dnneop{(p_n,p_m)\equiv \int \prod_{i=1}^6 dx_i e^{-N V(x_1,x_2,x_3,x_4,x_5,x_6)}
p_n(x_1) p_m(x_6) = h_n \delta_{n,m}}
Introducing again
the operators of multiplication by eigenvalues $Q_1$, $Q_2$, $Q_3$, 
$Q_4=Q_3^\dagger$, $Q_5=Q_2^\dagger$, $Q_6=Q_1^\dagger$ and the operator $P_1$
of derivation \wrt eigenvalues of $A_1$ 
(all regarded as acting on  $p_n(x_1)$), we find the master equations
\eqn\dnneme{\eqalign{
P_1 \over N &=Q_2-g Q_1^2 \cr
0&=Q_1-Q_3-g z_2 Q_2^2 \cr
0&=-Q_2+Q_3^\dagger-g z_1 Q_3^2 \cr}}
which determine the $h$'s completely.

As a direct consequence of \dnneme, the $Q_i$'s have a finite expansion in powers of
the shift operator $\sigma$ ($\sigma p_n=p_{n+1}$). The symmetry relation
\eqn\dnnesimV{ V(\omega x_1, \omega^2 x_2, \omega x_3,\omega^2 x_4, \omega x_5,\omega^2
x_6)=V(x_1,x_2,x_3,x_4,
x_5,x_6),
\qquad \omega=e^{2i\pi/3} }
analogous to the case of hard objects on bicolorable trivalent graphs, ensures
that the relation \sympn\ still holds for the orthogonal polynomials at hand.
Then, using relations
\dnneme\ and $\sigma^\dagger=\sigma^{-1} v$, where $v$ is still defined as the 
diagonal operator with entries $v_n=h_n/h_{n-1}$, we find the expansions
\eqn\qIexp{\eqalign{Q_1 & =\sigma + \sum_{k=1}^{11}\sigma^{-3k+1} s^{(k)}\cr 
Q_2 & = g \sigma^2 +\sum_{k=1}^6 \sigma^{-3k+2} t^{(k)}\cr
Q_3 & =-\sigma^4 g^3 z_2 + \sigma u^{(1)} + \sigma^{-2} u^{(2)} + \sigma^{-5}
u^{(3)} + \sigma^{-8} u^{(4)} \cr}}
where the $s^{(k)}$, $t^{(k)}$ and $u^{(k)}$ are diagonal operators in 
the $p_n$ basis. 
For simplicity we shall from now on go directly to the planar limit $n, N\to \infty$
as before with $x=n/N$ fixed, in which the operators $s^{(k)}$, 
$t^{(k)}$ and $u^{(k)}$ become functions of $x$,
while $\sigma$ now plays the role of a dummy scalar expansion parameter. 
The last two lines of \dnneme\ allow clearly to express the $s^{(k)}$
and $t^{(k)}$ in terms of the $u^{(k)}$.
Writing moreover the relation $Q_2=Q_3^\dagger -gz_1 Q_3^2$ at orders $8,5,2$ in $\sigma$ we 
may express $u^{(2)},u^{(3)},u^{(4)}$ as 
\eqn\utwothrefour{\eqalign{
u^{(2)}&={gv^2(1+z_1 (u^{(1)})^2)\over 1+2 g^4 z_1 z_2 v^2}\cr
u^{(3)}&= -2 g^4 z_1 z_2 u^{(1)} v^5 \cr
u^{(4)}&= g^7 z_1 z_2 ^2 v^8\cr}}
We finally get two equations determining $u^{(1)}$ and $v$ implicitly in terms
of $x$, by writing $Q_1=Q_3+g z_2 Q_2^2$ at order $1$ in $\sigma$
and $P_1/N=Q_2-g Q_1^2$ at order $-1$ in $\sigma$. 
Upon defining the rescaled quantities
\eqn\rescap{ \eqalign{
\alpha&={z_1\over z_2}\cr
V&= g^2 z_2 v\cr
U&= z_2 (u^{(1)})^2 \cr}}
we end up with the two equations 
\eqn\lutfin{\eqalign{
g^2 z_2^2 x &=\varphi(V,U)\equiv 
4 V^4 (1-2 \alpha^2 V^4) -2V^2 {(1+\alpha U)^2\over (1+2 \alpha V^2)^2} 
+V(1-20 \alpha^2 V^4) U \cr
z_2 &=\psi(V,U)\equiv U\left( 2V(1-4\alpha^2 V^4)
+{(1-2\alpha V^2-4 U \alpha^2 V^2)\over (1+2 \alpha V^2)}\right)^2 \cr}} 
This system generalizes \finmas\ in the sense that we must first
solve the second equation for $U(V)$ as an implicit function of $V$
(namely $z_2=\psi(V,U(V))$) and plug it back into the first equation 
to get the relation $g^2 z_2^2 x=\varphi(V)\equiv\varphi(V,U(V))$,
leading to the formula for the planar free energy through relation \thremof\  
(upon the substitution $g \to g^2 z_2$ in the denominator of the Log).
More precisely, the correct determination of $U$ is dictated by the
small $x$ limit in which $v\sim x$, hence $V\sim g^2 z_2 x$ and $U\to z_2$.

Before we turn to the general study of the critical lines of the
model, it is instructive to analyze the simple limiting cases
discussed in Sect.\ 3.1, namely $u^{-1}\to 0$ ($z_1\to 0$) for which 
we expect to recover the one-particle model of Sect.\ 2.3, and $(1/z_2)\to 0$ 
($z_1,z_2\to \infty$ with $u=z_2/z_1^2$ fixed) for which we expect
a first order transition.

For $z_1\to 0$, we simply take $\alpha =0$ in \lutfin\  
to write the second equation as $U=z_2/(1+2V)^2$ while the first
equation gives $g^2 z_2^2 x= V U -2 V^2(1-2 V)^2$. 
We therefore recover eqn.\ \finmas\ with $z_2\to z$. 

More interestingly, in the other limit, we must let the parameters scale
as $z_1=\hat{z}_1/\epsilon$,
$z_2=\hat{z}_2/\epsilon^2$, $g=\hat{g}\epsilon$ with 
$\epsilon\to 0$. From \rescap\ we deduce the 
other rescalings $\alpha=\hat{\alpha}\epsilon$,
$V=\hat{V}$ and $U=\hat{U}/\epsilon^2$. In this limit,
eqn.\ \lutfin\ becomes
\eqn\firsto{\eqalign{ \hat{g}^2\hat{z}_1^2x & = {1\over 2}
(2 \hat{U}\hat{V}\hat{\alpha}^2)
\big(1-(2 \hat{U}\hat{V}\hat{\alpha}^2)\big) \cr
\hat{z}_2 & =\hat{U}(2\hat{V}+1-4\hat{U}\hat{V}^2
\hat{\alpha}^2)^2\cr}}
The second line of \firsto\ may be recast as
\eqn\secundo{u^{-1}{1\over 2}\sqrt{{\hat{U}\over \hat{z}_2}}
\left(1-\sqrt{{\hat{U}\over \hat{z}_2}}\right)
= {1\over 2} (2 \hat{U}\hat{V}\hat{\alpha}^2)
\big(1-(2 \hat{U}\hat{V}\hat{\alpha}^2)\big)
}
with $u^{-1}=z_1^2/z_2=\hat{z}_1^2/\hat{z}_2$ as before.
This gives an alternative expression for 
$\hat{\varphi}(\hat{V})\equiv\hat{g}^2\hat{z}_1^2x$.
The maxima of $\hat{\varphi}(\hat{V})$ correspond clearly to either 
$2\hat{U}\hat{V}\hat{\alpha}^2=1/2$ or $\sqrt{\hat{U}/\hat{z}_2}=1/2$ 
leading respectively to the critical values $\hat{g}^2_1=1/(8\hat{z}_1^2)$ or 
$\hat{g}^2_2=u^{-1}/(8 \hat{z}_1^2)=1/(8 \hat{z}_2)$. As before, the choice 
of the correct determination
is best seen by considering the critical lines in the plane $(\hat{V},u^{-1})$.
Using \secundo\ with $2\hat{U}\hat{V}\hat{\alpha}^2=1/2$, we obtain the first
curve 
\eqn\solun{u^{-1}=
{\hat{z}_2\hat{V}\hat{\alpha}^2\over 2\sqrt{\hat{z}_2\hat{V}\hat{\alpha}^2}-1}}
Using \secundo\ with now $\sqrt{\hat{U}/\hat{z}_2}=1/2$, we get the second curve 
\eqn\soldeux{u^{-1}= \hat{z}_2\hat{V}\hat{\alpha}^2(2-\hat{z}_2\hat{V}
\hat{\alpha}^2)}
\fig{Critical lines in the 
$(\tilde{V}\equiv\hat{z}_2\hat{V}\hat{\alpha}^2,u^{-1})$ plane 
corresponding to $\hat{\varphi}'(\hat{V})=0$.
The correct line corresponds to the lowest value of $\tilde{V}$ and is
represented by a solid line. The first order transition point $u^{-1}=1$
is characterized by the contact between the two determinations of
$u^{-1}$ at which $\hat{g}^2$ jumps from $1/(8 \hat{z}_2)$ to
$1/(8 \hat{z}_1^2)$.}{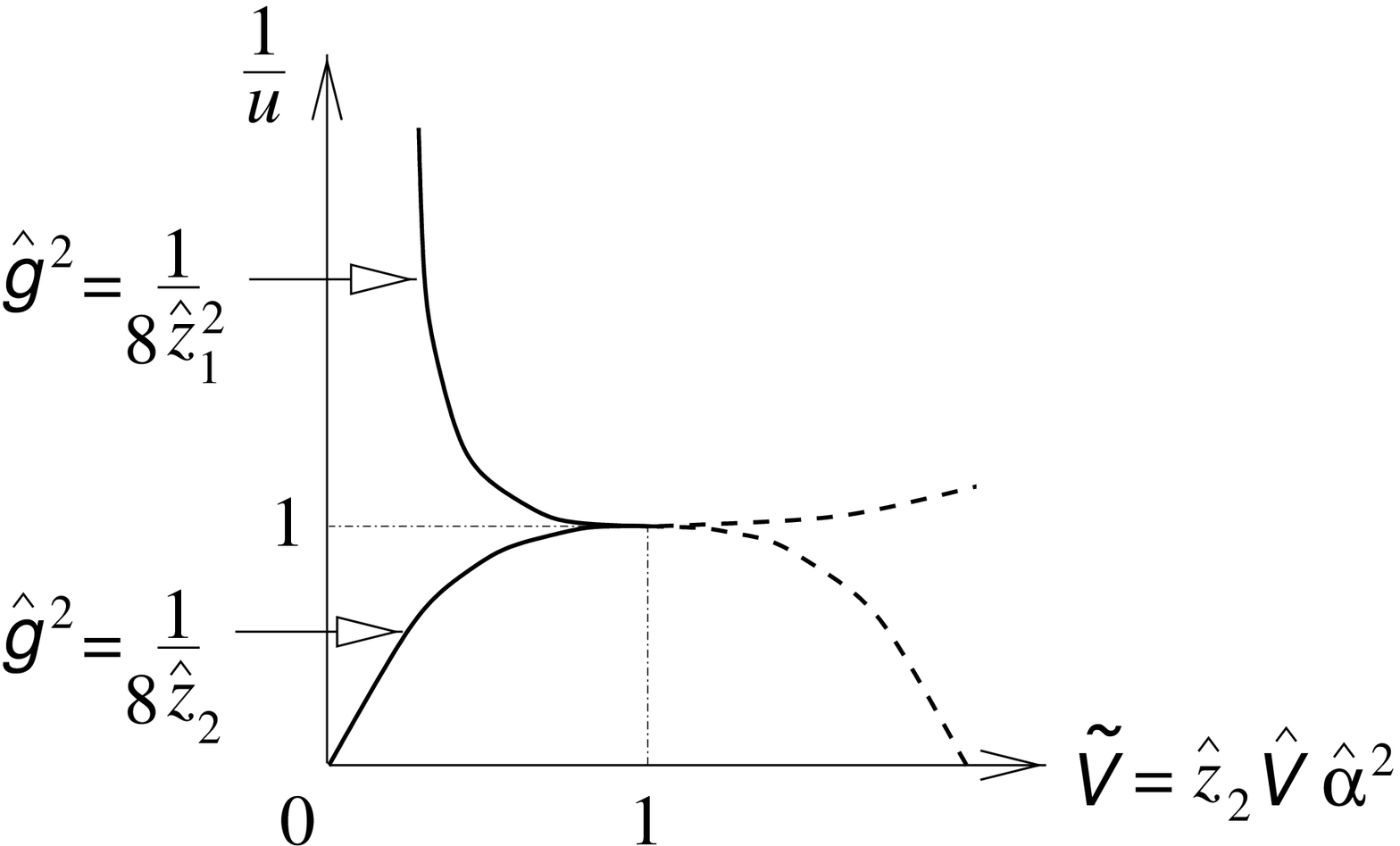}{12.truecm}
\figlabel\Vudiag
\fig{Critical line in the $(u^{-1},\hat{g}^2\hat{z}_1^2)$ plane
for the limiting case $z_1,z_2\to\infty$, $u^{-1}=z_1^2/z_2$ fixed. 
The behaviors of $\hat{\varphi}(\hat{V})$ below, at and above the 
transition are displayed in the small plots. The
slope of $\hat{g}^2\hat{z}_1^2$ is discontinuous at the first order
transition point $u^{-1}=1$.}{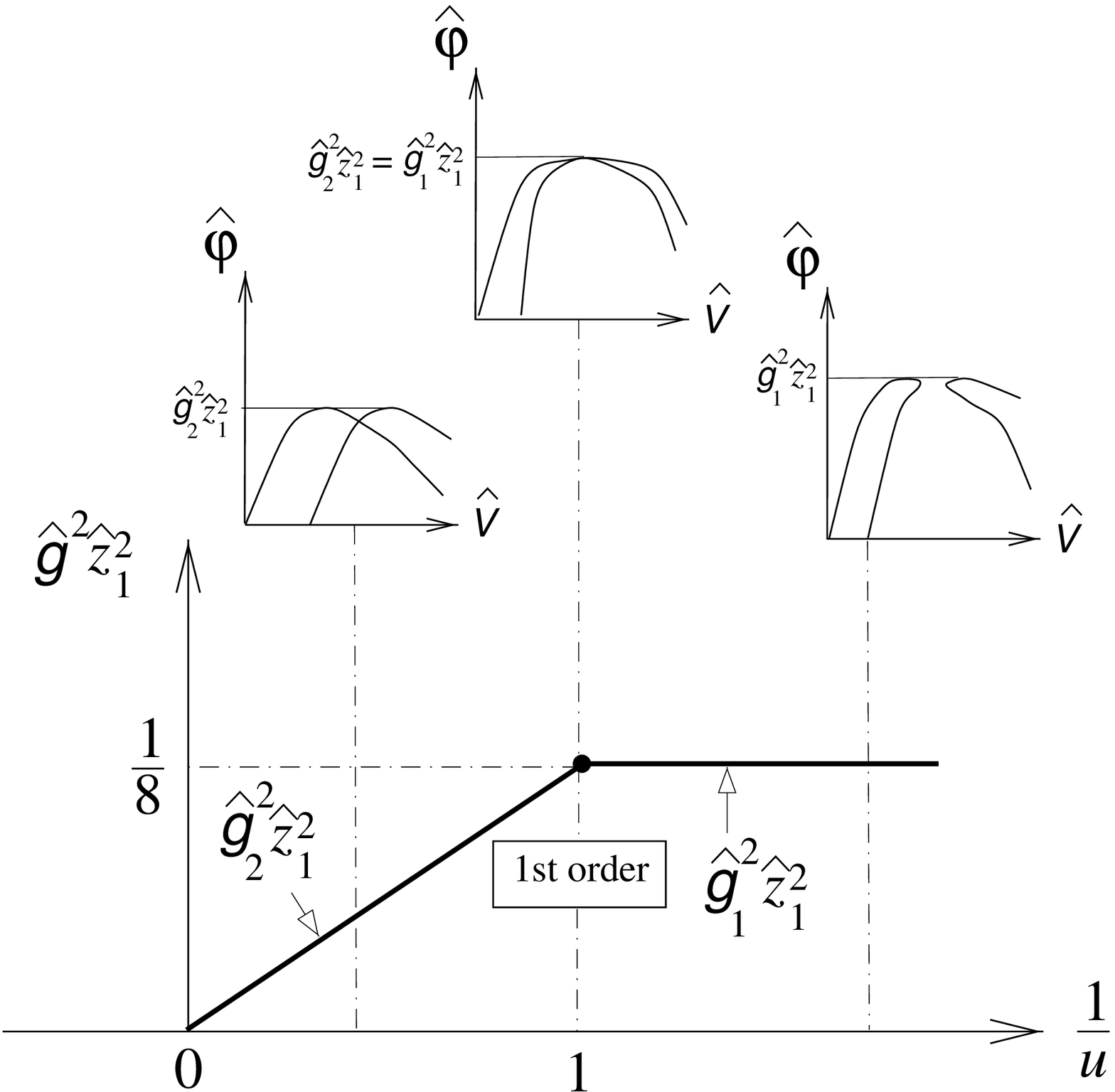}{12.truecm}
\figlabel\linefirst
\noindent The curves \solun\ and \soldeux\ are plotted in Fig.\ \Vudiag.
The solid portions correspond to the smallest values of $\hat{V}$
for fixed $u^{-1}$ which define the location of the relevant maxima of
$\hat{\varphi}(\hat{V})$ attained by the change of variables $x\to \hat{V}$.
The transition between the two curves takes place at $u=1$ where $\hat{g}^2$
changes expression from $\hat{g}_1^2$ to $\hat{g}_2^2$ (see Fig.\ \linefirst). 
This is clearly a first order transition as the slope of $\hat{g}^2(u)$ has a 
discontinuity at $u=1$. Note also that $\hat{\varphi}''({\hat V})$ is non-zero
at the transition point, therefore we have $\gamma=-1/2$ as in the case of
pure gravity.

Let us now turn to the general analysis of the complete phase diagram as obtained from
eqn.\ \lutfin. As before, we first look for critical lines characterized by
$\varphi'(V)=0=\partial_V\varphi-\partial_U\varphi (\partial_V\psi/\partial_U\psi)$.  
Using the explicit expressions for $\varphi(V,U)$ and $\psi(V,U)$, we immediately
get two possible conditions:
\eqn\ucrittwo{\eqalign{0&=(1-2\alpha V^2)^2(1-2V(1+2\alpha V^2)^2)+4 U \alpha^2V^2
(1+6\alpha V^2)\cr
0&=4V(1+2V(1+2\alpha V^2)^2)^2\cr
&\ \ \ -U(1-8\alpha V(1-V-15 \alpha V^3-52 \alpha^2 V^5
-50 \alpha^3 V^7))+4 U^2 \alpha^2 V\cr}}
which leads to the three determinations: $U=U_0$ solution of
the first line; $U=U_\pm$ conjugate solutions of the second line.
\fig{Critical lines in the $(V,z_2)$ plane for $z_2>0$ 
at some fixed (small enough)
value of $\alpha=z_1/z_2$, as obtained by setting $\varphi'(V)=0$. The
solid lines correspond to the correct branches given by the smallest
values of $V$. They correspond successively to the solutions $U_-$,
$U_0$ and $U_+$ of eqn.\ \ucrittwo. A continuous transition point
occurs at the crossing point $z_2=z_2^{(c)}$ of the two lowest
branches, with a situation analogous to that found in Fig.\ \Vzdiag\
at the point $z=z_+$. A discontinuous transition point occurs at
the contact point $z_2=z_2^{(d)}$ between the two upper branches, with
a situation analogous to that of Fig.\ \Vudiag\ at the point $u^{-1}=1$.
}{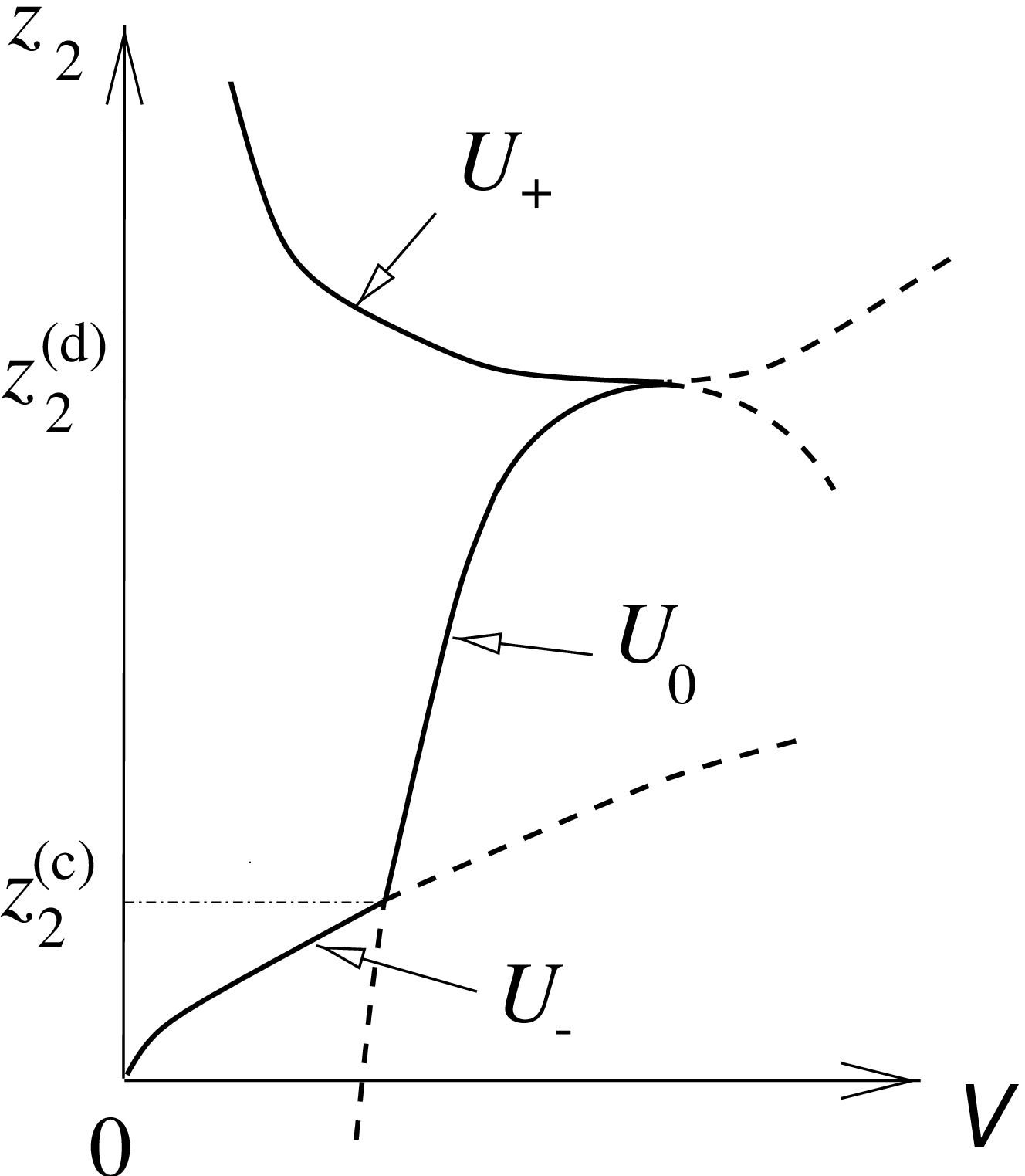}{8.truecm}
\figlabel\generic
When substituted into $z_2=\psi(V,U)$, this gives three branches
in the plane $(V,z_2)$ for fixed $\alpha$. 
Let us now restrict ourselves to $z_1,z_2>0$, hence $\alpha>0$.
For small enough positive values of $\alpha$ the branches in the
$(V,z_2)$ plane take the generic form displayed in Fig.\ \generic. 
As before, the correct
solution corresponds to the smallest value of $V$ for fixed $z_2$,
represented by solid lines in Fig.\ \generic.
We clearly identify two transitions at values $z_{2}^{(d)}$ and
$z_{2}^{(c)}$. Comparing the qualitative behavior of the curves in the
vicinity of the transition points with the behaviors obtained so far
in the two above limiting cases, we can identify
$z_2^{(d)}$ with a first order discontinuous transition point and 
$z_2^{(c)}$ with a continuous (critical Ising) transition point.

The transition points correspond as illustrated
in Fig.\ \generic\ to the coincidence of two branches. 
We can therefore obtain their location by expressing that
$U_0=U_+$ or $U_0=U_{-}$, which is done explicitly by writing that 
the solution $U_0$ of the first line of \ucrittwo\ also satisfies
the second line. We get the two possible conditions: 
\eqn\transloc{\eqalign{0&=1-V(1-2 V^2\alpha -20 (V^2\alpha)^2-
24 (V^2\alpha)^3)\cr 0&= 1-20 V^2\alpha +100 (V^2\alpha)^2
-2V (1-16 V^2\alpha -104 (V^2\alpha)^2 -448 (V^2\alpha)^3
+400 (V^2\alpha)^4)\cr}}
Introducing the variable $W=2 V^2 \alpha$, we can express all
the relevant quantities as rational fractions of $W$.
For the first line of \transloc\ we end up with
\eqn\firstpara{\eqalign{{1\over z_2}&= {W^2(1+W)^6(1-3W)^5\over 4(1-W)^6}\cr
{1\over u}&={(1-W)^6(1+W)^2\over 1-3W}\cr}}
For the second line of \transloc\ we get
\eqn\secupara{\eqalign{{1\over z_2}&= {(1-8W-26W^2-56W^3+25 W^4)^5\over
32(1-W)^6(1-5W)^4(1-8W-25W^2)^2}\cr
{1\over u}&={128W^2(1-W)^6(1-8W-25W^2)^2\over (1-5W)^4 
(1-8W-26W^2-56W^3+25 W^4)}\cr}}
It is easy to check that the points of the 
second curve \secupara\
correspond to crossings of branches such as that happening
at $z_2^{(c)}$ in Fig.\ \generic. The situation around these points
is totally analogous to that described on Fig.\ \linetri\ at
the Ising transition point, with in particular $\varphi''(V)=0$,
henceforth $\gamma=-1/3$. We identify this curve with a
line of continuous critical Ising transition points (CFT with $c=1/2$ 
coupled to 2D quantum gravity). 

The first curve \firstpara\ on the other hand corresponds
to a contact of branches such as that encountered at $z_2^{(d)}$.
The situation around this point is now similar to that found 
in Fig.\ \linefirst\ at the first order transition point.
On this line we have $\varphi''(V)\neq 0$, hence $\gamma=-1/2$ as
in the case of pure gravity (CFT with $c=0$ 
coupled to 2D quantum gravity).
However, the critical parameter $g_c^2$ has a discontinuity in its slope
across this line, hence so does the free energy.
We thus identify this curve with a line of first order transition
points. 
\fig{The phase diagram of the two-particle exclusion model on
random vertex bicolorable planar lattices in the $(1/z_2,u^{-1})$ plane 
for $z_2>0$,
as obtained from the exact solution of the six-matrix model \dnneint.
This phase diagram agrees with that of Fig.\ \phasediag\ 
with an Ising-like critical (solid) line meeting a first order
(dashed) one at a tricritical Ising point $(t)$. We have indicated 
the relevant branches
of $U$ solving eqn.\ \ucrittwo, namely $U_0$ in the ordered phase, and $U_\pm$
in the fluid one. We have also indicated a typical constant $\alpha=z_1/z_2$
hyperbola along which we encounter successively the two continuous
and discontinuous transitions of Fig.\ \generic.}{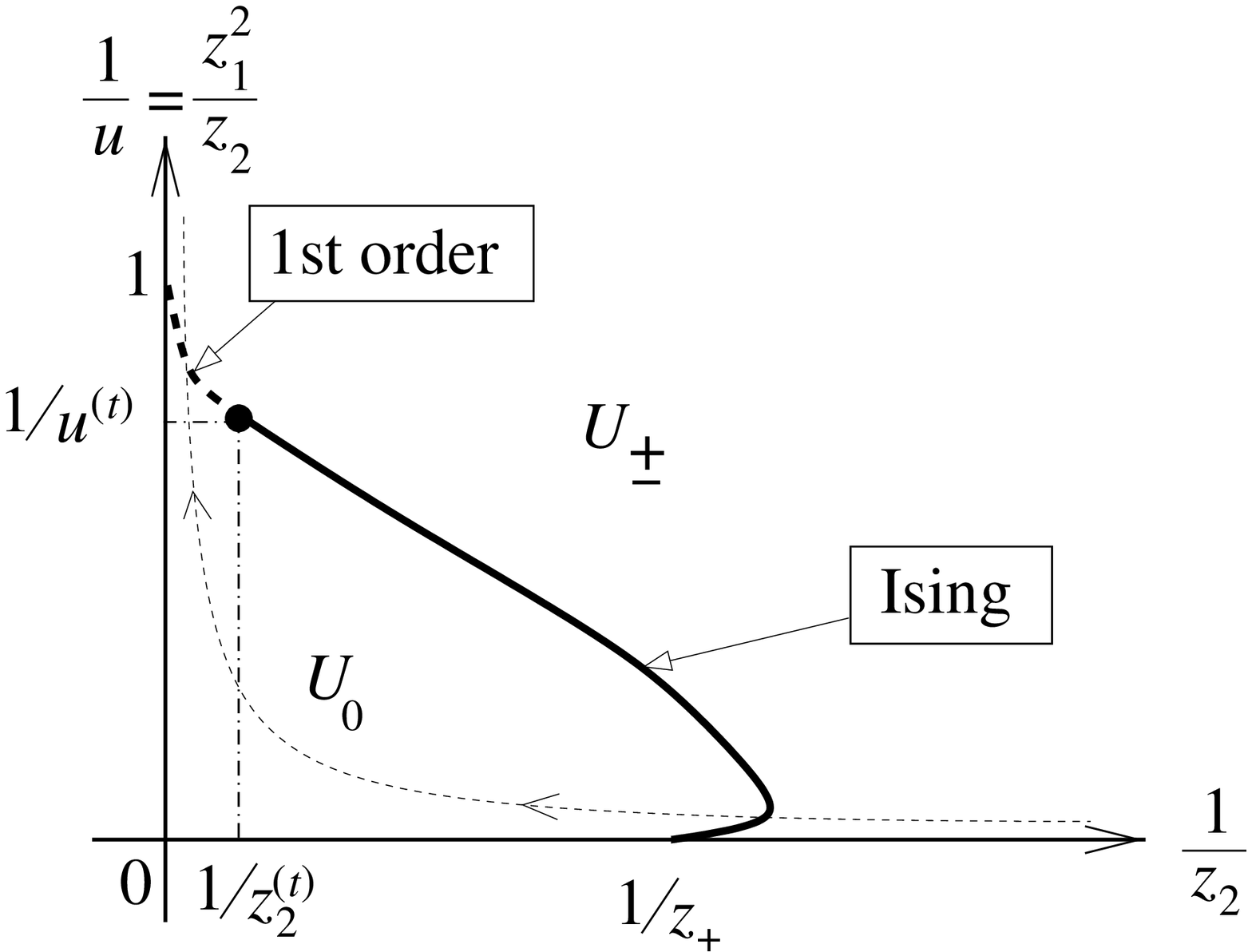}{13.truecm}
\figlabel\phases
These results are summarized in Fig.\ \phases\ 
where the first order transition line is represented by a dashed line
and the continuous one by a solid line. 
These two lines meet at a tricritical point with
\eqn\ztwocrit{\eqalign{ W^{(t)}&={\sqrt{41}-6 \over 5}\cr
{1\over z_2^{(t)}}&= {15633086927-2441464587\sqrt{41}\over 80000000}
=0.00107319\ldots\cr
{1\over u^{(t)}}&={ 32 (-564779+87849 \sqrt{41})\over 78125}=0.930177
\ldots\cr}}
Note that Fig.\ \generic\ corresponds to a typical small enough constant 
$\alpha$ section (hyperbola in the $(1/z_2,u^{-1})$ plane)
such as that represented in dotted line on Fig.\ \phases\ which 
crosses the two transition lines successively. The ordered phase lies 
below the transition lines and is described by the solution $U_0$
of the first line of \ucrittwo\ while the disordered phase lying
above the transition curves corresponds to $U_\pm$ solutions of
the second line of \ucrittwo. 

The above phase diagram gives an {\it explicit} realization of that described
qualitatively in Sect.\ 3.1. To complete our study, let us now show
that the tricritical point \ztwocrit\ displays the expected behavior 
for a tricritical Ising transition point (CFT with $c(4,5)=7/10$) 
coupled to gravity. A first evidence comes from the fact that the
string susceptibility exponent at this point is $\gamma=-1/4$ as
$\varphi'=\varphi''=\varphi'''=0$ at this point while 
$\varphi^{(4)}\neq 0$. Note that the vanishing of $\varphi''$ holds 
generically for all points of the critical Ising line \secupara, 
as one readily checks by direct calculation. The vanishing of
$\varphi'''$ holds only at the tricritical point, as may be checked by
a direct calculation too. The tricritical
Ising CFT is the only unitary CFT with $\gamma=-1/4$ when coupled 
to gravity\foot{$\gamma=-2/(p+q-1)$ for a central charge 
$c(p,q)$ hence $p+q=9$ and $q-p=1$ by unitarity yield $p=4$, $q=5$.}. 
A second check can be performed by computing the thermal exponent
$\alpha$. More precisely, one can define two thermal exponents
pertaining to two thermal operators with conformal dimensions
$h_{33}=1/10$ and $h_{32}=3/5$, and the corresponding dressed
dimensions when coupled to gravity $\Delta_{33}=1/4$ 
and $\Delta_{32}=3/4$. As explained in Appendix A,
the most relevant one ($\Phi_{33}$) 
governs the generic approach to the critical point through
$f\sim (z_2-z_2^{(t)})^{2-\alpha}$ with 
$\alpha=(1-2\Delta_{33})/(1-\Delta_{33})=2/3$,
while the other operator ($\Phi_{32}$) governs the fine-tuned approach along
a line tangent to the critical curves \firstpara\ and \secupara,
with a behavior $f\sim (z_2-z_2^{(t)})^{2-\alpha'}$ with 
$\alpha'=(1-2\Delta_{32})/(1-\Delta_{32})=-2$.
To obtain the value of the first exponent in our model, a simple 
procedure consists in first fixing the ratio
$z_1/z_2$, expanding $U$ for both lines of \ucrittwo\ in
terms of $V-V^{(t)}$ and substituting the result into \lutfin.
We finally get $g_c^2-(g^{(t)})^2 = a (V-V^{(t)})^3 +b (V-V^{(t)})^4 +\ldots$
and $z_2-z_2^{(t)} = a' (V-V^{(t)})^3 +b' (V-V^{(t)})^4 +\ldots$ which
upon inversion leads to $g_c^2-(g^{(t)})^2= a''(z_2-z_2^{(t)})
+b''(z_2-z_2^{(t)})^{4/3}$ with values of $a''$ and $b''\neq 0$ independent
of the determination of $U$, hence $2-\alpha=4/3$ as expected.
To get the second exponent, we may approach the tricritical point
by traveling along the transition lines \firstpara\ and \secupara.
For the first transition line we have
\eqn\gcfirst{g_c^2= {W^2 (1+W)^4(1-3W)^6(1-2W+3W^2+20 W^3+3 W^4-50 W^5-35 W^6)
\over 32 (1-W)^{12}} }
For the second transition line we have
\eqn\gcsecu{g_c^2= {15(2-12 W-41 W^2+54 W^3-19W^4)(1-8W-26W^2-56W^3+25W^4)^6\over
8192(1-W)^{12}(1-5W)^2 (1-8W-25W^2)^4} }
Using the corresponding parametric values of $z_2$ 
\firstpara\ and \secupara\ respectively, we easily check that
the first three derivatives of $g_c^2$ \wrt $z_2$ match at the critical point
$W=W^{(t)}$, while the fourth one is different. This discontinuity
corresponds to $2-\alpha'=4$, hence $\alpha'=-2$.

\fig{Critical line in the $(V,z_2)$ plane for $z_2<0$ as obtained by setting 
$\varphi'(V)=0$ for fixed (small enough in modulus) $\alpha=z_1/z_2$. As usual, 
the correct portion of the curve corresponds to the lowest 
value of $\vert V\vert$ and is represented by a solid line. The Lee-Yang 
critical point $z_2^-$ is characterized by $dz_2/dV=0$ and corresponds to the 
merging and annihilation
of the two extrema $1$ and $2$ in (a) with $\varphi''(V)=0$. 
By increasing $|\alpha |$, we reach the situation (b) where 
a third extremum $3$ merges with them. This defines the tricritical 
point $z_2^{(t')}$ satisfying in addition $\varphi'''(V)=0$.}{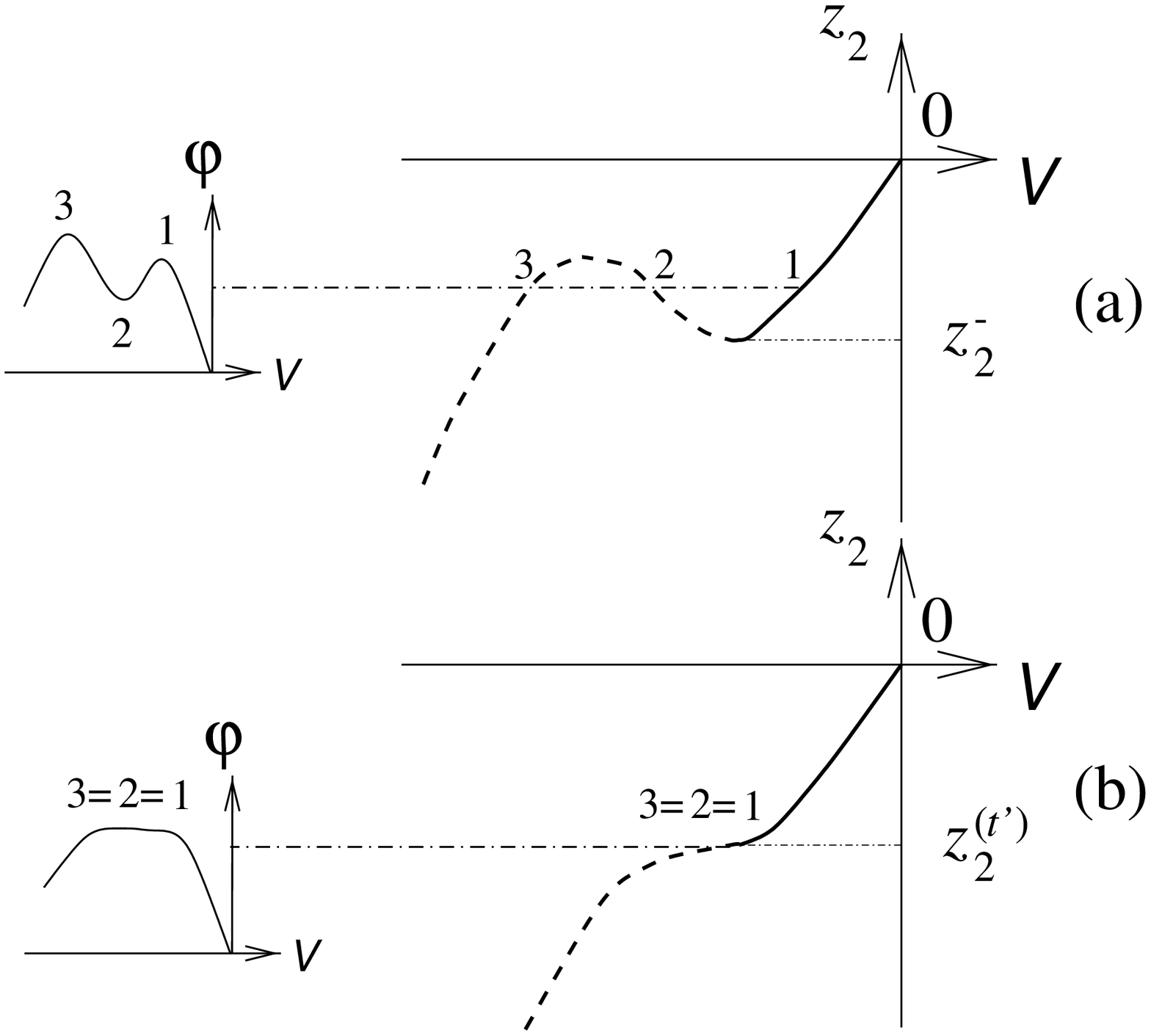}{9.cm}
\figlabel\neg

For completeness let us finally discuss the case $z_1>0$ and $z_2<0$ ($u^{-1}<0$).
Looking again for the critical lines with $\varphi'=0$, 
the relevant branch for $U$ yielding the physical determination
of $\varphi$ is given by the second line of \ucrittwo\ i.e. $U=U_-$, say. 
Once substituted back into the second line of \lutfin, this gives a critical line
in the $(V,z_2)$ plane, such as that plotted in Fig.\ \neg\ (a) (for $\alpha=z_1/z_2$
small enough).  We recover the Lee-Yang singularity point at some $z_2^{-}(u)$,
characterized by $\varphi''(V)=0$ ($\gamma=-1/3$)
due to the merging of the maximum of $\varphi$
with its further minimum. 
Eliminating $U$, we finally obtain the following parametric curve
\eqn\paraneq{\eqalign{ {1\over z_2}&= -{(5+100W+326W^2+820W^3-2275W^4)^5\over
512 (1-5W)^6 (1+35W)^4(1+20W+35W^2)^2}\cr
{1\over u}&=-{2048 W^2  (1-5W)^6 (1+20W+35W^2)^2\over
(1+35W)^4(5+100W+326W^2+820W^3-2275W^4)}\cr}}
with the parameter $W=-\alpha V^2\geq 0$.
In Fig.\ \neg\ (a),
we note the existence of a further maximum of $\varphi$ 
never attained by the change of variables $x\to V$ except when these three
extrema merge simultaneously (see Fig.\ \neg\ (b)). This point corresponds
to a higher order multicritical point with $\varphi'''=0$, hence
with a string susceptibility exponent $\gamma=-1/4$. This tricritical point
corresponds to the values 
\eqn\threneg{\eqalign{
W^{(t')}&= {8\sqrt{14}-21\over 455} \cr
{1\over z_2^{(t')}}&= -{172647361044 \sqrt{14}+645414154777\over 
317007031250}= -4.07373\ldots \cr 
{1\over u^{(t')}}&= -{524288(2401452\sqrt{14}-8699159)\over 10274243531825}
=-0.0146072\ldots \cr}}
We thus conclude that the line of Lee-Yang critical points $z_2^{-}(u)$ ends
at this higher critical point $z_2^{(t')}$. We identify this point with the only 
non-unitary CFT with $\gamma=-2/(p+q-1)=-1/4$, i.e. $c(2,7)=-68/7$, coupled to 2D
quantum gravity.  
A confirmation of this fact may be obtained by computing the thermal
exponents of the theory $\alpha$ and $\alpha'$ characterizing the singularity
of the free energy (or equivalently of $g_c^2$) as $z_2$ approaches the
tricritical value $z_2^{(t')}$. Like in the unitary case,
the exponent $\alpha$ governs the generic approach to this point in
the $(1/z_2,u^{-1})$ plane, while $\alpha'$ governs the fine-tuned approach along the
line of Lee-Yang singular points. As shown in Appendix A, the predicted values
are $\alpha=2/3$ and $\alpha'=1/2$. As before, we obtain the first exponent
in our model by
expanding $g^2$ and $z_2$ along a curve with, say, $z_1/z_2=$const. We find
generically 
$g_c^2-(g^{(t')})^2 = a (V-V^{(t')})^3 +b (V-V^{(t')})^4 +\ldots$
and $z_2-z_2^{(t')} = a' (V-V^{(t')})^3 +b' (V-V^{(t')})^4 +\ldots$ which
upon inversion leads to $g_c^2-(g^{(t')})^2= a''(z_2-z_2^{(t')})
+b''(z_2-z_2^{(t')})^{4/3}$ with $b''\neq 0$, which gives $2-\alpha=4/3$.
To compute the second exponent, we use the parametric equations   
\paraneq\ and the corresponding value of $g_c^2$
\eqn\valgc{ g_c^2={15(5+100W+326W^2+820W^3-2275W^4)^6 (2-60W+151
W^2+630W^3-2275W^4)\over 2097152 (1-5W)^{12}(1+35W)^2(1+20W+35W^2)^4} }
and expand $g_c^2$ and $z_2$ around the tricritical point $W^{(t')}$,
with
$g_c^2-(g^{(t')})^2 = a (W-W^{(t')})^2 +b (W-W^{(t')})^3 +\ldots$
and $z_2-z_2^{(t')} = a' (W-W^{(t')})^2 +b' (W-W^{(t')})^3 +\ldots$ which
upon inversion leads to $g_c^2-(g^{(t')})^2= a''(z_2-z_2^{(t')})
+b''(z_2-z_2^{(t')}))^{3/2}$ with $b''\neq 0$, which gives $2-\alpha'=3/2$.

\fig{The phase diagram of the two-particle exclusion model on
random vertex bicolorable planar lattices in the $(1/z_2,u^{-1})$ plane
for $z_2<0$,
as obtained from the exact solution of the six-matrix model \dnneint.
A (solid) line
of Lee-Yang type critical points terminates at a tricritical point $(t')$
where it meets another first order (dashed) transition line. These two
lines form the border of the region of the phase diagram
where an oscillatory behavior is observed in the canonical partition
function $Z_A(z_1,z_2)$.}{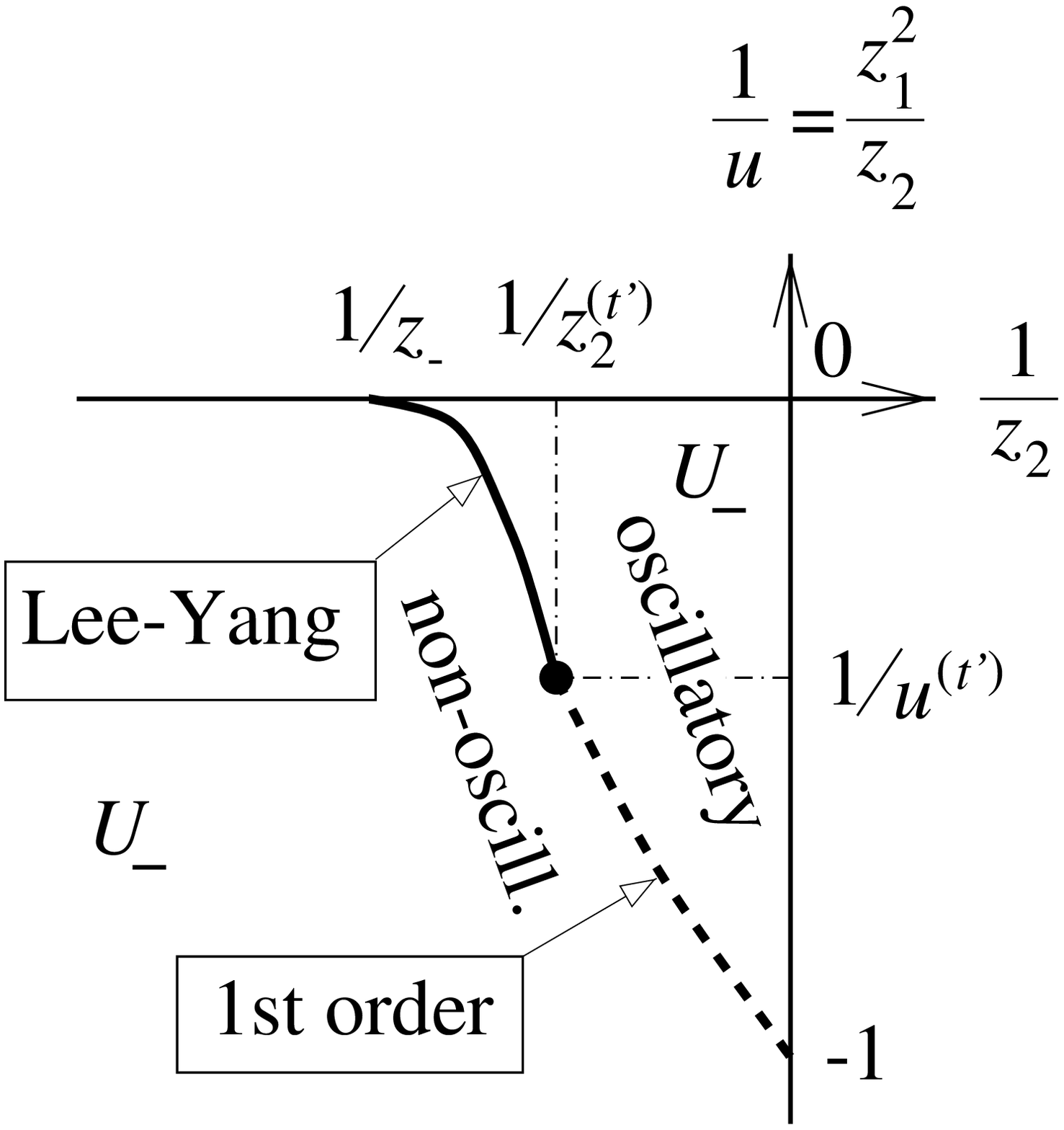}{9.truecm}
\figlabel\phaseneg

As discussed at the end of Sect.\ 2.2, the Lee-Yang critical line separates
a small $\vert z_2\vert $ region where $g_c$ is real from a large
$\vert z_2\vert $ region where it becomes complex and generates an
oscillatory behavior of the form \oscillo\ for the canonical 
partition function $Z_A(z_1,z_2)$ 
for planar graphs of fixed area $A$. This separation between oscillatory
and non-oscillatory behaviors extends beyond the tricritical point
$(t')$ in the form of a first order transition line as depicted
on Fig.\ \phaseneg. This line corresponds to a situation where the
two complex conjugate values of $g$ corresponding to the complex
conjugate solutions to $\varphi'=0$ cross in modulus the real value
of $g$ corresponding to the real solution to $\varphi'=0$.
The transition through this line is first order in the sense that
the thermodynamic free energy has a discontinuous slope across the line.  
The line clearly originates at the tricritical point $(t')$ where the
three values of $g$ are real and coincide. More interestingly,
it is easy to see that it terminates at the point $(1/z_2,u^{-1})=(0,-1)$.
Indeed, along the axis $1/z_2=0$, we have to compare the real value
$\hat{g}_1=1/\sqrt{8 \hat{z_1}^2}$ and the two complex conjugate values
$\hat{g}_2=\pm i/\sqrt{8 \vert\hat{z}_2\vert}$. For 
$u^{-1}>-1$, we have an oscillatory behavior of the partition
function $\hat{Z}_A(\hat{z}_1,\hat{z}_2)=\lim \epsilon^A Z_A(z_1,z_2)
\sim (-8\hat{z}_2)^{A/2}$, while for $u^{-1}<-1$ we simply have 
a non-oscillatory behavior 
$\hat{Z}_A(\hat{z}_1,\hat{z}_2)\sim (2\sqrt{2}\hat{z}_1)^{A}$.

\newsec{Conclusion and Discussion}

In this paper, we have shown how to reach critical and multicritical points
in the context of nearest neighbor exclusion models. We have displayed a number
of exact solutions for various models defined on random planar lattices. 
The crucial outcome of this analysis is the importance of the colorability
condition of the underlying lattice. Indeed, we have shown that critical
points based on a $\IZ_2$ symmetry such as the critical Ising and tricritical
Ising models are reproduced for exclusion models under the condition that the
lattice itself be vertex-bicolorable. 
In view of understanding the physics of the corresponding models on regular lattices, 
the outcome of the random lattice solutions is twofold:
first it sorts out which features of the lattice itself take a relevant part in
the models' critical behavior; secondly it allows for an exact solution in the planar
limit that clearly identifies the critical universality classes for regular lattice models
(such as the celebrated Hard Square model) which are still unsolved directly. 
More precisely, this study allows to infer that the bipartite nature of
the square and honeycomb lattices indeed translates into a critical
Ising-like universality class for their crystallization transition point.

We have shown how to extend the definition of hard particle models to reach
higher order critical points, by introducing a two-particle exclusion model
in which each site may be empty, singly or doubly occupied with an exclusion constraint
that a total of at most two particles may share the same edge. A straightforward
generalization consists in having a $k$-particle exclusion model, where 
sites can be occupied by $0$, $1$, $2$, ... up to $k$ particles, with a weight
$z_i$ for an occupancy by $i$ particles, and with the exclusion
constraint that a total of at most $k$ particles may share the same edge.     
When defined on a regular lattice, the exclusion constraint is easily turned into
an non-overlapping constraint for appropriate tiles of various sizes.
When defined on a random vertex-bicolorable lattice, these models are
amenable to a $(2k+2)$ matrix integral, with potential
\eqn\potgenk{ V(A_1,A_2,...,A_{2k+2})=\sum_{i=1}^{2k+1} (-1)^{i+1} A_i A_{i+1}
-{g \over 3} \sum_{i=0}^k z_i (A_{2i+1}^3+A_{2(k-i)+2}^3) }
for trivalent graphs, with a weight $g$ per vertex (and $z_0=1$).
Remarkably enough, the exclusion rule translates into a chain-like quadratic interaction
between the matrices, which makes the models exactly solvable by means
of standard orthogonal polynomial techniques. 
With the model \potgenk, we expect to be able to reach multicritical points 
governed by CFTs with central charges $c(p,q)$ with $p+q=2k+3$, by appropriately 
fine-tuning the $z_i$'s, in order to reach the highest possible critical string
susceptibility exponent $\gamma=-1/(k+2)$. We also expect the situation to be identical
on vertex-bicolorable regular lattices.

One may also address the question of the behavior of the same particle exclusion models
on lattices with other colorability properties. From the exact solution of the Hard Hexagon
model on the triangular (hence vertex-tricolorable) lattice, it is natural to expect that 
the above exclusion models, when defined on vertex-tricolorable (fixed or random) lattices,
give rise to critical and multicritical three-state Potts models. Unfortunately,
the corresponding matrix models are no longer directly solvable by means of orthogonal
polynomials.

More generally, the vertex-$k$-colorability of the lattice is likely to generate
within the context of exclusion models critical
points with $\IZ_k$ symmetry such as the $k$-state Potts model or 
the $\IZ_k$ model of Fateev and Zamolodchikov \FZ. 
We know however that for large enough $k>4$ we lose
the continuous transition in the Potts case, while in the other case the central charge
$c=2(k-1)/(k+2)>1$ forbids any meaningful coupling to 2D quantum gravity. 

Note finally that the above study provides an example of critical
phenomena whose coupling to gravity is sensitive to the type of
graphs summed over. While the precise connectivity (tri- or
tetra-valency) of the lattices is unimportant, their Eulerian 
(bipartite) or Euclidean (arbitrary) character is relevant.
This situation is reminiscent of that of fully-packed loop
models [\xref\GKN,\xref\DGK].
\bigskip
\noindent{\bf Acknowledgments} We thank T. Garel for a useful 
discussion on tricritical points and G. Akemann for a critical
reading of the manuscript. J. B. is supported by the \'Ecole 
normale sup\'erieure.

\appendix{A}{String susceptibility and thermal 
exponents for minimal CFT coupled to 2D quantum gravity}

The matrix models in general give a discretized version of matter systems
coupled to 2D quantum gravity, in the form of discrete statistical models
defined on random graphs accounting for the fluctuations of space-time.
In the continuum approach to 2D quantum gravity, one may relate the
properties of some critical matter system in fixed geometry (typically a CFT with
central charge $c$ in the plane) to that of the same matter system
coupled to 2D quantum gravity. The precise connection involves
the celebrated KPZ formula \KPZ\ expressing the string
susceptibility exponent for a unitary CFT of central charge $c$ coupled to 
2D quantum gravity:
\eqn\gamkpz{ \gamma(c)={c-1-\sqrt{(1-c)(25-c)} \over 12}}
as well as that the dimension $\Delta$ of the gravitational
dressing $\phi$ of a primary operator of the CFT with dimension $h$:
\eqn\deltkpz{ \Delta(h,c)={\sqrt{1-c+24 h} -\sqrt{1-c} \over \sqrt{25-c} -\sqrt{1-c}}}
This dimension measures the singular behavior of the corresponding 
two-point correlator
as $\langle \phi \phi \rangle \sim (\mu_c-\mu)^{2\Delta(h,c) -\gamma(c)}$, 
when the cosmological constant $\mu$ approaches its critical value $\mu_c$. 
The above formulae simplify drastically when considering minimal CFT with
central charge
\eqn\cpq{ c(p,q)=1 -6 {(p-q)^2 \over pq} }
with say $q\geq p+1$ and $p\wedge q=1$, 
and operator spectrum
\eqn\spepq{ h_{r,s}= {(qr-ps)^2 -(p-q)^2\over 4 p q} }
with $qr-ps>0$, $1\leq r\leq p-1$, $1\leq s\leq q-1$.
Eqns.\ \gamkpz\ and \deltkpz\ reduce in this case to
\eqn\graviqp{ \gamma(c(p,q))=1 - {q \over p} \qquad
\Delta_{r,s}\equiv\Delta\big(h_{r,s},c(p,q)\big)= {(r-1) q-(s-1)p \over 2 p} }

For a unitary theory ($q=p+1$), the most relevant operator of the theory is the
identity with $h_{1,1}=0$ and its gravitational dressing known as the ``puncture operator"
$P$ has $\Delta_{1,1}=0$ as well. This operator measures  the area of the random
surface, which allows to identify the deviation $\mu_c-\mu$ with $g_c-g$ 
in the corresponding matrix model, and to interpret $\gamma=-1/p$ as the
exponent governing the singularity of the free energy.
In the case of pure gravity ($p=2,q=3$), we have $\gamma=-1/2$, and the only operator
is $P$. In the case of the critical Ising model ($p=3,q=4$) we have $\gamma=-1/3$
and three operators: the puncture $P$, the dressed energy operator $\Phi_{2,1}$
and the dressed spin operator $\Phi_{2,2}$, with respectively $\Delta_{1,1}=0$,
$\Delta_{2,1}=2/3$ and $\Delta_{2,2}=1/6$. 
The dressed energy is the thermal operator, with coupling $z_c-z\sim
(\mu_c-\mu)^{1-\Delta_{2,1}}=(g_c-g)^{1/3}$, hence upon inversion we get a singularity
of the free energy of the form given by $g_c-g\sim (z_c-z)^{2-\alpha}$,
with $\alpha=-1$. 
In the case of the tricritical Ising model ($p=4,q=5$), we have $\gamma=-1/4$ and
there are six operators, among which we distinguish the puncture $P$ and
the two thermal operators $\Phi_{3,3}$ and $\Phi_{3,2}$, with respectively
$\Delta_{3,3}=1/4$ and $\Delta_{3,2}=3/4$. The generic thermal perturbation of the model
is governed by the most relevant operator $\Phi_{3,3}$ with coupling
$z_2-z_2^{(t)}\sim (\mu_c-\mu)^{1-\Delta_{3,3}}=(g_c-g)^{3/4}$, yielding the
singularity of the free energy $g_c-g\sim (z_2-z_2^{(t)})^{2-\alpha}$ with $\alpha=2/3$.
By fine-tuning the parameters one may approach the tricritical point on a line
along which the contribution of $\Phi_{3,3}$ is cancelled, and therefore the next
most relevant thermal operator $\Phi_{3,2}$ takes over. This leads analogously
to another thermal exponent $\alpha'$ with $2-\alpha'=1/(1-\Delta_{3,2})$,
hence $\alpha'=-2$.

For non-unitary theories with 
$q>p+1$, the above formulae must be
interpreted carefully to account for the fact that the identity is no longer the most
relevant operator. Indeed, the scale of the deviation from criticality
is set instead by the operator of smallest (negative)
dimension $h_0=(1-(p-q)^2)/(4pq)$ (with $qr-ps=1$) corresponding to the 
gravitationally dressed operator
$\Phi_0$ with dimension $\Delta_0=(p-q+1)/(2p)$. 
The latter operator has the coupling $(\mu_c-\mu)^{1-\Delta_0}$
which allows to identify the 
deviation from criticality $(g_c-g)$ in the matrix model
as $g_c-g=(\mu_c-\mu)^{1-\Delta_0}$. Note that this general relation also holds
in the unitary case $q=p+1$, where it reduces to $g_c-g=\mu_c-\mu$, 
as $h_0=\Delta_0=0$.
The most singular part of the free energy is due to the presence of $\Phi_0$
and may be obtained by writing that $d^2 f/dg^2\vert_{\rm sing}\sim
\langle\Phi_0\Phi_0\rangle$. 
We can therefore write
\eqn\behatwoptzero{\langle \Phi_0 \Phi_0 \rangle 
\sim (\mu_c-\mu)^{2 \Delta_0-\gamma(c(p,q))}  
\sim (g_c -g)^{-\gamma} }
from which we deduce the corrected string susceptibility exponent 
$\gamma$ of the matrix model: 
\eqn\relatgam{ -\gamma= {2 \Delta_0-\gamma(c(p,q))\over 1-\Delta_0} ={2 \over p+q-1} }
To compute the thermal exponent we need to identify the next most relevant thermal
operator say $\Phi_1$ with dimension $\Delta_1$ with coupling proportional 
to $(\mu_c-\mu)^{1-\Delta_1}\sim (g_c-g)^{1/(2-\alpha)}$, leading to
$2-\alpha=(1-\Delta_0)/(1-\Delta_1)$.
 
For the Lee-Yang edge singularity, we have $p=2,q=5$, 
and therefore $\Delta_0=\Delta_{1,2}=-1/2$
and $\gamma=-1/3$. 
In this case, the deviation from the critical ``temperature" $z_c-z$
is coupled to the  next most relevant operator of the theory,
which turns out to be the puncture operator $\Phi_1=P$ with $\Delta_1=0$, 
leading to $\alpha=1+\Delta_0=1/2$.  
For the case $p=2,q=7$ of Sect.\ 3.2, we have $\Delta_0=\Delta_{1,3}=-1$ and $\gamma=-1/4$.
A generic deviation from criticality $z_2^{(t')}- z_2$
is coupled to the operator $\Phi_1\equiv \Phi_{1,2}$ with $\Delta_1=\Delta_{1,2}=-1/2$, hence
yields a thermal exponent $\alpha=2/3$. In a fine-tuned approach to the critical point,
we may cancel the contribution of $\Phi_0$, in which case $\Phi_1$ now plays the
role of the most relevant operator, with $\Delta_0'=\Delta_1$, while the thermal
operator becomes $P$, with $\Delta_1'=0$. We deduce the fine-tuned thermal exponent
$\alpha'=1+\Delta_0'=1+\Delta_1=1/2$ identical to that of the Lee-Yang case.

\appendix{B}{Hard particles on a random tetravalent lattice: double scaling limit}

In the following, we derive the double scaling limit of the model of hard
particles on a random lattice (non-necessarily vertex-bicolorable). We show that
the renormalized string susceptibility obeys the Lee-Yang differential equation.
We first rewrite the eqns \durdur\ and \durmou\ in components, namely
\eqn\sysonehs{\eqalign{ {n \over N}&=v_n -r_n
-gz(s_n+s_{n+1}+s_{n+2}+r_n(r_{n-1}+r_n+r_{n+1}) \cr
{\tilde r}_{n} &= v_{n}(1+gz(r_{n-1}+r_{n}+r_{n+1}))\cr
{\tilde s}_{n}&=gz v_{n}v_{n-1}v_{n-2}\cr}}
and
\eqn\systwohs{\eqalign{ {n \over N}&=v_{n}-g({\tilde s}_{n}+
{\tilde s}_{n+1}+{\tilde s}_{n+2}+{\tilde r}_{n}({\tilde r}_{n-1}+
{\tilde r}_{n}+{\tilde r}_{n+1})\cr
r_n &= gv_n({\tilde r}_{n-1}+{\tilde r}_n+{\tilde r}_{n+1})\cr
s_n &= g v_n v_{n-1}v_{n-2}\cr}}
Introducing the new coefficients
\eqn\newcoef{ R_n=g r_n, \ S_n=g^2 s_n, \ {\tilde R}_n=g {\tilde r}_n, \
{\tilde S}_n=g^2{\tilde s}_n, \ V_n=g {\tilde v}_n }
we simply get
\eqn\simplhs{\eqalign{
{\tilde R}_n&=V_n(1+z(R_{n-1}+R_n+R_{n+1}))\cr
{\tilde S}_n&=z V_n V_{n-1}V_{n-2}\cr
R_n&=V_n({\tilde R}_{n-1}+{\tilde R}_n+{\tilde R}_{n+1})\cr
S_n&=V_n V_{n-1}V_{n-2}\cr
g{n \over N}&=V_n -{R_n{\tilde R}_n\over V_n}-zV_n(V_{n-1}V_{n-2}
+V_{n-1}V_{n+1}+V_{n+1}V_{n+2}) \cr}}
where both first lines of \sysonehs\ and \systwohs\ turn out to be equivalent
to the last line of \simplhs.
When $N$ becomes large, all sequences tend to smooth functions of $x=n/N$,
and setting $a=1/N$,
we now make the following scaling ansatz on $V_n\equiv V(x)=V(1-a^2 v(x))$,
for some unknown function $v(x)$ for which we will derive a differential equation.
We must first solve the first and third lines of
\simplhs\ for $r(x)$ order by order in $a$,
where $R_n\equiv R(x)=R(1-a^2 r(x))$, and where the values of $V,R,z$ are taken along
the critical line \parameq, namely with
\eqn\eqpara{ z={12 u(1+3u)^2\over (1-3u)^8}, \ \ 
V={(1-3u)^4\over 6(1+3u)}, \ \ R= {(1-3u)^7\over 12(1+3u)^2}} 
with the result
\eqn\ressol{\eqalign{ r(x)&={2 v(x) \over 1-3u}+a^2 {1+9u\over 3(1-3u)^2}(v''(x)-3v(x)^2)
-{a^4 \over 36(1-3u)^3}((1+54u+117u^2) v^{(4)}(x)\cr
&-12(1+42u+81u^2)v(x)v''(x) -288u(1+3u)v'(x)^2
+432 u(1+3u)v(x)^3)+O(a^6)\cr}}
We now expand the last line of \simplhs\ up to order $6$ in $a$, after setting $u$ to its critical
value $u_c=(2\sqrt{5}-5)/15$, with the final result
\eqn\findiflee{ {g_c-g x\over g_c}=a^6(v(x)^3-{1\over 2} v'(x)^2-v(x)v''(x)+{1\over
10}v^{(4)}(x))}
Upon introducing the renormalized cosmological constant $y=(g_c-gx)/(a^{6/7}g_c)$
and appropriately rescaling $v(x)\to a^{-12/7}v(y)$, we finally
get the standard differential equation for the renormalized string susceptibility $v(y)$
\eqn\difyl{y=v(y)^3-{1\over 2} v'(y)^2-v(y)v''(y)+{1\over 10}v^{(4)}(y)} 
which is easily identified with that of the Lee-Yang edge singularity coupled to 2D quantum
gravity \DGZ.

\appendix{C}{Hard particles on arbitrary random trivalent lattices}

In the following we show that the trivalent lattice version of \hsmod\
leads to the same qualitative physics, namely a unique critical point
in the universality class of the Lee-Yang edge singularity coupled to
2D quantum gravity.
 
We start with the trivalent version of the matrix model \hsmod:
\eqn\htmod{\eqalign{
Z^{(3)}_N(g,z)&=\int dA dB e^{-N{\rm Tr}\, V(A,B)} \cr
V(A,B)&=-{1\over 2}A^2 +AB -g{B^3\over 3} -gz {A^3\over 3}\cr}}
where $A,B$ are Hermitian with size $N\times N$, and the measure
is normalized so that $Z^{(3)}_N(0,0)=1$.
Comparing with \hsmod, the bi-orthogonal polynomials are no longer even/odd as
$V$ is no longer even. We may still write the trivalent version of \opsys:
\eqn\newopsys{\eqalign{{P_1\over N}&= Q_2^\dagger -Q_1 -gz Q_1^2\cr
{P_2\over N}&=Q_1^\dagger-g Q_2^2\cr}}
and show that the $Q$'s have finite range, with
\eqn\acttrQ{\eqalign{
Q_1 &= \sigma+r+\sigma^{-1} s+\sigma^{-2} t\cr
Q_2 &= \tau+ {\tilde r}+\tau^{-1}{\tilde s}+\tau^{-2}{\tilde t}\cr}}
where $\sigma,\tau$ denote the shift operators acting respectively on the
left and right bi-orthogonal polynomials.
In components, the equations \newopsys\ now read
\eqn\sysonehstr{\eqalign{ {n \over N}&=v_n -s_n
-gz(t_n+t_{n+1}+s_n(r_{n-1}+r_n) \cr
{\tilde r}_{n}&= (r_n+gz(s_n+s_{n+1}+r_{n}^2))\cr
{\tilde s}_{n}&=v_{n}(1+gz(r_n+r_{n-1}))\cr
{\tilde t}_{n}&=g z v_{n}v_{n-1}\cr}}
and
\eqn\systwohstr{\eqalign{
{n \over N}&=v_{n}-g({\tilde t}_{n}+
{\tilde t}_{n+1}+{\tilde s}_{n}({\tilde r}_n+{\tilde r}_{n-1}))\cr
r_n&= g ({\tilde s}_n+{\tilde s}_{n+1}+{\tilde r}_{n}^2)\cr
s_n&= g v_n({\tilde r}_{n-1}+{\tilde r}_n))\cr
t_n&= g v_n v_{n-1}\cr}}
For large $n,N$ with $x=n/N$, we get algebraic equations for
$V,R,S,T(x)$ respectively limits of $g^2z v_n,gz r_n,g^2z s_n, g^3z t_n$
and their tilded counterparts:
\eqn\countertr{\eqalign{
g^2z x &=  V-S-2 zT-2RS\cr
{\tilde R}&= R(1+R) +2z S\cr
{\tilde S}&= V(1+2R)\cr
{\tilde T}&= V^2\cr
R&= 2 {\tilde S} +{\tilde R}^2/z\cr
S&= 2 V {\tilde R}/z\cr
T&=V^2/z\cr}}
easily solved in the form of an algebraic equation for $R$ as a function of $V$
\eqn\algrvtr{ R\left(z- {R(1+R)^2\over (1-4V)^3}\right)={2zV\over 1-4V} }
and a master equation for the dependence on $x$:
\eqn\masttr{ g^2z^2 x\equiv \varphi(V)=zV(1-2V)-2{VR(1+R)(1+2R)\over 1-4V} }
Writing that $\varphi'=\varphi''=0$, and eliminating $z$ and $V$, we are left
with a sixth order equation for the critical value of $R$, namely
\eqn\criR{13+266 R+1810 R^2+5920 R^3 +10200 R^4 +8748 R^5+ 2916 R^6=0}
only one root of which leads to a positive value of $g^2z^2$ through \masttr\
at $x=1$. This leads to a unique critical point at
\eqn\crititr{\eqalign{R_c&=-0.090430\ldots\qquad 
V_c=-0.036173\ldots \cr
z_c&=-0.16565\ldots \qquad g_c=\sqrt{\varphi(V_c)}/|z_c|=0.28105\ldots\cr} }
with the critical exponent $\gamma_{str}=-1/3$, and the critical point is
in the (non-unitary) class of the Yang-Lee edge singularity as in the tetravalent case.
It is easy to check that the scaling ansatz $V_n=V(1-a^2v(x))$, $a=1/N$, still leads,
upon solving \sysonehstr\-\systwohstr\ order by order in $a$, to the same
differential equation \findiflee\ in which $g\to g^2z^2$.

\appendix{D}{Hard particles on trivalent bicolorable graphs: double scaling limit}

In the following, 
we complete the identification of both tricritical points
\tripoint\ for the model \htmodfour\ by
deriving the corresponding differential
equations for the renormalized version of the string susceptibility $V(x)$.
To derive the double-scaling limit of the matrix model \htmodfour,
let us first write in components the complete equations \internat:
\eqn\getfromeq{\eqalign{
s_n^{(0)}&=g\cr
s_{n}^{(2)}&=-g^3 z  v_{n}v_{n-1}v_{n-2}v_{n-3} \cr
v_{n}&=s_{n}^{(1)}+g^2 z v_{n} (s_{n-1}^{(1)}+s_{n+1}^{(1)} )\cr
r_n^{(1)}&= g v_{n-1}v_n +g^2 z(s_n^{(2)}+s_{n+2}^{(2)})+g z s_n^{(1)}s_{n-1}^{(1)}\cr
{n\over N}&= s_n^{(1)} - g (r_n^{(1)}+r_{n+1}^{(1)}) \cr}}
Setting $V_n= g^2 z v_n$, $S_n=g^2 z s_n^{(1)}$, we finally get
\eqn\finhtr{\eqalign{
0&= S_n- V_n(1-S_{n-1}-S_{n+1}) \cr
g^2 z^2{n\over N} &= z S_n(1-S_{n-1}-S_{n+1})-V_n(V_{n-1}+V_{n+1})\cr
&+V_{n-1}V_{n}V_{n+1}(V_{n+2}+V_{n-2})
+V_n(V_{n-1}V_{n-2}V_{n-3}+V_{n+1}V_{n+2}V_{n+3})\cr}}
We make the following scaling ansatz
$V(x)=V(1-a^2 v(x))$,
$a=1/N$ a small parameter.
We first solve the equation $V_n(1-S_{n-1}-S_{n+1})=S_n$ order by order in $a$,
with the result
\eqn\orderby{\eqalign{ S(x)&=S(1-a^2 s(x))\cr
s(x)&={v(x)\over 1+2V}-a^2 {V \over (1+2V)^2}v''(x)
-a^2 {2 V\over (1+2 V)^2} v(x)^2
-{a^4\over 12} {V(1-10V)\over (1+2V)^3} v^{(4)}(x) \cr
&+a^4{4V^2\over (1+2V)^3}v'(x)^2
-a^4{V(1-4V)\over (1+2V)^2} v(x)v''(x)+a^4 {4 V^2\over (1+2V)^3} v(x)^3+
O(a^6)\cr}}
which we then substitute into the second line of \finhtr\ and Taylor-expand in $a$.
Apart from the term of order zero, $\varphi(V)=g^2t^2$, the
lowest order terms are in $a^6$.
We then simply get at leading order
\eqn\diffeq{\eqalign{
{g_i^2t_i^2-g^2t^2 x\over g_i^2 t_i^2}&= a^6 ( A_i v(x)^3 + B_i v(x)v''(x)
+C_i v'(x)^2+D_iv^{(4)}(x))+O(a^8)\cr
A_+&={8\over 15}, \qquad A_-={4\over 3}\cr
B_+&=-{6\over 5}, \qquad B_-=-{2}\cr
C_+&=-{3\over 5}, \qquad C_-=-1\cr
D_+&={1\over 5}, \qquad D_-={3\over 10}\cr}}
where the index $i=\pm$ refers to the critical point $z_\pm$.
Upon setting
$y=(g_i^2t_i^2-g^2t^2 x)/(g_i^2 t_i^2 a^{6/7})$
and rescaling respectively $v(x)\to 9v(y)/(4a^{12/7})$ and 
$v(x) \to 3v(y)/(2a^{12/7})$, the differential
equations take the standard form
\eqn\standeq{\eqalign{
(1)\ \ v^3-vv''-{1\over 2} (v')^2+{2\over 27} v^{(4)} &= y\cr
(2)\ \ v^3-vv''-{1\over 2} (v')^2+{1\over 10} v^{(4)} &= y\cr}}
which we immediately identify
with the differential equations governing the double scaling limit of
respectively the Ising model and the Lee-Yang edge singularity \DGZ.

\appendix{E}{Hard particles on tetravalent bicolorable random graphs}

The study is quite analogous to that of Sect.\ 2.3.
The matrix integral takes the same form as \htmodfour, but with the potential
\eqn\potfour{ V(A_1,A_2,A_3,A_4)=A_1 A_2 -A_2 A_3+A_3 A_4-{g \over 4}(A_1^4+A_4^4)
-{g z\over 4} (A_2^4+A_3^4) }
We still have the symmetry $A_i\leftrightarrow A_{5-i}$ for $i=1,2$, hence introducing
again a family of monic orthogonal polynomials $p_n$, and
keeping notations as n \derto, we get the two equations
\eqn\mastfour{\eqalign{ {P_1\over N}&= Q_2 -g Q_1^3 \cr
0 &= Q_1-Q_2^\dagger -gz Q_2^3 \cr}}
The potential now satisfies the following additional symmetry property, replacing \simV:
\eqn\symfour{ V(ix_1,-ix_2,ix_3,-ix_4)=V(x_1,x_2,x_3,x_4), \qquad i^2=-1 }
The operators $Q_1,Q_2$ still have finite range, as a consequence
of \mastfour, and read, thanks to the symmetry \symfour:
\eqn\graq{\eqalign{
Q_1&=\sigma+ \sum_{j=1}^7\sigma^{1-4j} r^{(j)} \cr
Q_2&=g \sigma^3+\sigma^{-1} s^{(1)}+\sigma^{-5} s^{(2)}+\sigma^{-9}s^{(3)}\cr}}
with the usual shift operator $\sigma$, with $\sigma^\dagger=\sigma^{-1}v$,
where $v_n=h_n/h_{n-1}$ as usual, $h_n=(p_n,p_n)$ the square norm of $p_n$.
The adjoint of $Q_2$ reads
\eqn\adjfour{ Q_2^\dagger=g (\sigma^{-1}v)^3+s^{(1)}v^{-1}\sigma+s^{(2)}(v^{-1}\sigma)^5
+s^{(3)} (v^{-1}\sigma)^9}
and the equations \mastfour\ boil down to
\eqn\boil{\eqalign{
s^{(3)}&=-g^4 z \sigma^9 (\sigma^{-1} v)^9 \cr
s^{(2)}&=-g^3 z (\sigma^5 s^{(1)}+\sigma^2 s^{(1)}\sigma^3
+\sigma^{-1}s^{(1)}\sigma^6)(\sigma^{-1} v)^5\cr
s^{(1)}&=v -g z(\sigma
s^{(2)}\sigma^{-1}+\sigma^{-2}s^{(2)}\sigma^2+\sigma^{-5}s^{(2)}\sigma^5)\cr
&-g^2 z ( \sigma^2(s^{(1)}\sigma^{-1})^2+\sigma^{-1}s^{(1)}\sigma^2s^{(1)}\sigma^{-1} 
+\sigma^{-2}(s^{(1)}\sigma)^2)\cr
r^{(1)}&=g \sigma^3(\sigma^{-1} v)^3+gz(s^{(3)}+\sigma^{-3} s^{(3)}\sigma^{3}+
\sigma^{-6} s^{(3)}\sigma^{6})\cr
&+g^2z (\sigma s^{(2)}\sigma^{-1}s^{(1)}+\sigma^5\sigma^{(1)}\sigma^{-5}s^{(2)}+
\sigma^{-2}s^{(2)}\sigma^2s^{(1)}+\sigma^2s^{(1)}\sigma^{-2}s^{(2)}\cr
&+\sigma^{-3}s^{(1)}\sigma s^{(2)}\sigma^2+\sigma^2 s^{(1)}\sigma^{-5}s^{(2)}\sigma^3)
+gz \sigma^3(s^{(1)}\sigma^{-1})^3\cr
{\nu\over N}&= s^{(1)}-g(r^{(1)}+\sigma^{-1}r^{(1)}\sigma+\sigma^{-2}r^{(1)}\sigma^2) \cr}}
or equivalently in components:
\eqn\implic{\eqalign{
s_n^{(3)}&= -g^4 z v_n v_{n-1}...v_{n-8}\cr
s_n^{(2)}&=- g^3 z v_n v_{n-1}...v_{n-4}(s_{n-5}^{(1)}+s_{n-2}^{(1)}+s_{n+1}^{(1)})\cr
s_n^{(1)}&=v_n-gz\big( g^2(s_{n-1}^{(2)}+s_{n+2}^{(2)}+s_{n+5}^{(2)})
+g(s_{n-1}^{(1)}s_{n-2}^{(1)}+s_{n-1}^{(1)}s_{n+1}^{(1)}
+s_{n+1}^{(1)}s_{n+2}^{(1)})\big)\cr
r_n^{(1)}&=gv_nv_{n-1}v_{n-2}+gz\big( g^2(s_{n}^{(3)}+s_{n+3}^{(3)}+s_{n+6}^{(3)})
+g(s_n^{(1)}s_{n-1}^{(2)}+s_n^{(2)}s_{n-5}^{(1)}+s_n^{(1)}s_{n+2}^{(2)}\cr
&+s_n^{(2)}s_{n-2}^{(1)}+s_{n+3}^{(1)}s_{n+2}^{(2)}
+s_{n+3}^{(2)}s_{n-2}^{(1)})+s_n^{(1)}s_{n-1}^{(1)}s_{n-2}^{(1)}\big)\cr
{n\over N}&= s_n^{(1)}- g(r_n^{(1)}+r_{n+1}^{(1)}+r_{n+2}^{(1)}) \cr}}
Upon the redefinitions
\eqn\redefour{ V_n=g v_n, \qquad
R_n=g^2r_n^{(1)}, \qquad S_n=g s_n^{(1)}, \qquad T_n=g^2s_n^{(2)},
\qquad U_n=g^3s_n^{(3)} }
these equations finally reduce to
\eqn\maineqs{\eqalign{
U_n&=-zV_nV_{n-1}...V_{n-8}\cr
T_n&=-zV_nV_{n-1}...V_{n-4}(S_{n-5}+S_{n-2}+S_{n+1})\cr
S_n&=V_n(1-z(T_{n-1}+T_{n+2}+T_{n+5}+S_nS_{n-1}+S_{n-1}S_{n+1}+S_{n+1}S_{n+2}))\cr
R_n&=V_nV_{n-1}V_{n-2}+z(U_n+U_{n+3}+U_{n+6}\cr
&+S_nT_{n-1}+T_nS_{n-5}+S_nT_{n+2}+T_nS_{n-2}
+S_{n+3}T_{n+2}+T_{n+3}S_{n-2}+S_nS_{n-1}S_{n-2})\cr
g {n\over N}&= S_n-(R_n+R_{n+1}+R_{n+2}) \cr}}
In the planar limit, each sequence tends to a function of $x=n/N$, $n,N\to \infty$,
which we label by the same capital letter. Namely writing
$U=-zV^9$, $T=-3zV^5S$, and also introducing $\Sigma=S/V$ we find
\eqn\planarfour{\eqalign{
1&=\Sigma(1-9z^2V^6)+3zV^2 \Sigma^2 \cr
gx&\equiv\varphi(V)=\Sigma V-3(V^3+zV^3\Sigma^3-18z^2V^7\Sigma^2-3z^2V^9)\cr
&=-3V^3(1-3z^2V^6)+V\Sigma^2 (1+45z^2V^6)\cr}}
where we have used the first equation to simplify the second. The critical line
is the solution of $\varphi'(V)=0$.
The first line of \planarfour\ allows to compute
\eqn\dsv{ {d\Sigma\over dV}={6z V\Sigma(9zV^4-\Sigma) \over 1+6zV^2\Sigma-9z^2V^6}}
Setting $W=zV^3$, we find
\eqn\phipfour{ \varphi'(V)=-{((1-5W)\Sigma-3V(1+W))((1+5W)\Sigma+3V(1-W))
(2W\Sigma-V(1-W^2)) \over (2W\Sigma+V(1-W^2)) } }
hence we have three solutions
\eqn\trisol{\eqalign{ (1)\ \ \Sigma&={3V(1+W)\over 1-5W} \cr
(2)\ \ \Sigma&=-{3V(1-W)\over 1+5W}\cr
(3)\ \ \Sigma&={V(1-W^2)\over 2W}\cr}}
The first line of \planarfour\ allows to express $V$ in terms of $W$:
\eqn\valV{\eqalign{ (1)\ \ V&={1\over 3(1-3W+5W^2)}\left({1-5W\over 1+W}\right)^2\cr
(2)\ \ V&=-{1\over 3(1+3W+5W^2)}\left({1+5W\over 1-W}\right)^2\cr
(3)\ \ V&={4W\over 3(1-W^2)^2}\cr}}
We deduce the value of $z={W\over 3V^3}$:
\eqn\valz{\eqalign{
(1)\ \ z&={9W(1+W)^6(1-3W+5W^2)^3\over (1-5W)^6}\cr
(2)\ \ z&=-{9W(1-W)^6(1+3W+5W^2)^3\over (1+5W)^6}\cr
(3)\ \ z&={9(1-W^2)^6\over 64 W^2}\cr}}
and finally that of $g=\varphi(V)$:
\eqn\valgfour{\eqalign{
(1)\ \ g&={2(1-5W)^4(3+24W-10W^2+40W^3+35W^4)\over 27(1+W)^6(1-3W+5W^2)^3}\cr
(2)\ \ g&={2(1+5W)^4(3-24W+10W^2-40W^3+35W^4)\over 27(1-W)^6(1+3W+5W^2)^3}\cr
(3)\ \ g&={16W(1+W^2)(1-10W+5W^2)\over 27 (1-W^2)^6}\cr}}
The tricritical points are solution in addition of $\varphi''(V)=0$.
In the three above case, we get
\eqn\tricfour{\varphi''(V)=\left\{ \matrix{{18 V(1-12W+5W^2)(-1-20W+35W^2)\over (1-5W)^2}
& {\rm in} \ {\rm case}\ (1)\cr
{18 V(1+12W+5W^2)(-1+20W+35W^2)\over (1+5W)^2} & {\rm in} \ {\rm case}\ (2)\cr
-{3V\over 16 W^2}(1+5W^2)(1-12W+5W^2)(1+12W+5W^2)  & {\rm in} \ {\rm case}\ (3)\cr}
\right.}
As in the trivalent case of Sect.\ 2.3, each of the lines (1) and (2)
have two tricritical points, one of which is a cusp, the other coming from the
intersection with the critical curve (3). Moreover the curve (1) is always
reached before (2). The cusp solves $1+20W-35W^2=0$, while the intersection
solves $1-12W+5W^2=0$. Again, one of the two branches of these equations
is always reached before the other. We end up with a qualitative picture
identical to that of Fig.\linetri, with the cusp and intersection
corresponding respectively to the values $W_{\pm}$:
\eqn\valwpm{ W_-= {10-3\sqrt{15}\over 35},\qquad W_+={6-\sqrt{31}\over 5} }
while the tricritical point corresponds to the values $(z_\pm,g_\pm)$
given by the case (1) in \valgfour\ and \valz,
respectively with $W=W_\pm$ of \valwpm, with the exact values
\eqn\exagz{\eqalign{
z_-&=-{15683(-83151+26080\sqrt{15})\over 2573571875}=-.136568...\cr
g_-&={2000(129105-24881\sqrt{15})\over 235782657}=.277724...\cr
z_+&={6561(146327-21472\sqrt{31})\over 9765625}=17.989334...\cr
g_+&={80(14903-1067\sqrt{31})\over 14348907}=.049967...\cr}}

To get the double scaling limit, we set
\eqn\setdoub{ V_n=V(1-a^2 v(x)), \qquad S_n=\Sigma V(1-a^2s(x)) }
and first solve the third equation of \maineqs\ order by order in $a$, with the
result
\eqn\svalfour{\eqalign{
s(x)&= {1-5W\over 1+W} v(x)+a^2 {W(1-5W)(3-10W+5W^2)\over
(1+W)^2(1+5W^2)}(3v^2(x)-2v''(x))\cr
&-a^4 {W(1-5W)(3-155W+650W^2-1530W^3+1375W^4-175W^5)\over (1+W)^3(1+5 W^2)} v(x)^3\cr
&+a^4 {W(1-5W)(3-205W+355W^2-85W^3)\over (1+W)^3(1+5W^2)} v'(x)^2\cr
&+a^4 {12W(1-5W)(1-29W+70W^2-190W^3+225W^4-25W^5)\over (1+W)^3(1+5W^2)^2}v(x)v''(x)\cr
&-a^4 {W(1-5W)(9-445W+721W^2-121W^3)\over (1+W)^3(1+5W^2)} v^{(4)}(x)+O(a^6) \cr}}
We then substitute this and \setdoub\ into the last line of \maineqs, and Taylor-expand
in $a$.
The final result reads
\eqn\finfour{\eqalign{
{g_i-gx\over g_i}&=a^6(A_i v(x)^3 +B_i v(x)v''(x)
+C_i v'(x)^2+D_i v^{(4)}(x))+O(a^8)\cr
A_-&={5\over 2}D_- \qquad A_+={3\over 2}D_+ \cr
B_-&=-{5}D_- \qquad B_+=-{9\over 2}D_+ \cr
C_-&=-{5\over 2}D_- \qquad C_+=-{9\over 4}D_+ \cr
D_-&={27(5-3\sqrt{15})\over 110} \qquad D_+={291-49 \sqrt{31}\over 5125} \cr}}
which may be put back in the standard forms \standeq. We conclude that the situation
for the hard particle model on vertex-bicolorable 
tetravalent random lattices is qualitatively the
same as that for trivalent ones: we find two critical points at $z_-=-.136568...$ and
$z_+=17.989334...$ respectively in the universality classes
of the Lee-Yang and Ising critical models on random lattices. 
Note that this model is precisely
the gravitational version of the classical Hard Square model, as, in the dual picture
we are considering non-overlapping square tiles on random, face-bicolorable planar 
quadrangulations (dual to planar tetravalent vertex-bicolorable random lattices),
hence a random version of the square lattice incorporating its vertex-bicolorability.

\listrefs

\end